\documentclass[]{aa}
\usepackage{epsfig}
\usepackage{natbib}
\bibpunct{(}{)}{;}{a}{}{,}

\newcommand{\Fbol}{F\ensuremath{_\mathrm{bol}}}
\newcommand{\FeoH}{\ensuremath{\left[\mathrm{Fe}/\mathrm{H}\right]}}
\newcommand{\MoH}{\ensuremath{\left[\mathrm{M}/\mathrm{H}\right]}}
\newcommand{\thetaUD}{\ensuremath{\theta_\mathrm{UD}}}
\newcommand{\thetaR}{\ensuremath{\theta_\mathrm{R}}}

\begin{document}

\title{Broad-band photometric colors and effective temperature calibrations for late-type giants. I. Z=0.02}

\author { A. Ku\v{c}inskas \inst{1,2,3}, P.H. Hauschildt \inst{4}, H.-G. Ludwig \inst{1}, I. Brott \inst{4,5}, \\
        V. Vansevi\v{c}ius \inst{6}, L. Lindegren \inst{1}, T. Tanab\'{e} \inst{7}, F. Allard \inst{8}}
\institute{
           Lund Observatory, Lund University, Box 43, SE-221 00, Lund, Sweden
    \and   National Astronomical Observatory of Japan, Mitaka, Tokyo, 181-8588, Japan
    \and   Institute of Theoretical Physics and Astronomy, Go\v {s}tauto 12, Vilnius 01108, Lithuania
    \and   Hamburger Sternwarte, Gojenbergsweg 112, 21029 Hamburg, Germany
    \and   INTEGRAL Science Data Centre, Chemin d'Ecogia 16, 1290 Versoix, Switzerland
    \and   Institute of Physics, Savanoriu 231, Vilnius 02300, Lithuania
    \and   Institute of Astronomy, The University of Tokyo, Mitaka, Tokyo, 181-0015, Japan
    \and   Centre de Recherche Astronomique de Lyon, \'{E}cole Normale Sup\'erieure, Lyon, Cedex 07, 69364 France
           }

\offprints{A. Ku\v{c}inskas, \email{ak@itpa.lt}}

\date{Received 9 March 2005 / Accepted 23 June 2005}

\titlerunning{Broad-band photometric colors and effective temperature calibrations for late-type giants. I.}

\authorrunning{Ku\v{c}inskas et al.}

\abstract{We present new synthetic broad-band photometric colors
for late-type giants based on synthetic spectra calculated with
the {\tt PHOENIX} model atmosphere code. The grid covers effective
temperatures $T_{\rm eff}=3000\dots 5000$\,K, gravities $\log
g=-0.5\dots{+3.5}$, and metallicities ${\rm
[M/H]}=+0.5\dots{-4.0}$. We show that individual broad-band
photometric colors are strongly affected by model parameters such
as molecular opacities, gravity, microturbulent velocity, and
stellar mass. Our exploratory 3D modeling of a prototypical
late-type giant shows that convection has a noticeable effect on
the photometric colors too, as it alters significantly both the
vertical and horizontal thermal structures in the outer
atmosphere. The differences between colors calculated with full 3D
hydrodynamical and 1D model atmospheres are significant (e.g.,
$\Delta(V-K)\sim0.2$\,mag), translating into offsets in effective
temperature of up to $\sim$70\,K. For a sample of 74 late-type
giants in the Solar neighborhood, with interferometric effective
temperatures and broad-band photometry available in the
literature, we compare observed colors with a new {\tt PHOENIX}
grid of synthetic photometric colors, as well as with photometric
colors calculated with the {\tt MARCS} and {\tt ATLAS} model
atmosphere codes. We find good agreement of the new synthetic
colors with observations and published $T_{\rm eff}$--color and
color--color relations, especially in the $T_{\rm eff}$--$(V-K)$,
$T_{\rm eff}$--$(J-K)$ and $(J-K)$--$(V-K)$ planes. Deviations
from the observed trends in the $T_{\rm eff}$--color planes are
generally within $\pm100$\,K for $T_{\rm eff}=3500$ to 4800\,K.
Synthetic colors calculated with different stellar atmosphere
models agree to $\pm100$\,K, within a large range of effective
temperatures and gravities. The comparison of the observed and
synthetic spectra of late-type giants shows that discrepancies
result from the differences both in the strengths of various
spectral lines/bands (especially those of molecular bands, such as
TiO, H$_2$O, CO) and the continuum level. Finally, we derive
several new $T_{\rm eff}$--$\log g$--color relations for late-type
giants at solar-metallicity (valid for $T_{\rm eff}=3500$ to
4800\,K), based both on the observed effective temperatures and
colors of the nearby giants, and synthetic colors produced with
{\tt PHOENIX}, {\tt MARCS} and {\tt ATLAS} model atmospheres.
\thanks{Table~2 is available in electronic form at the CDS via anonymous ftp to cdsarc.u-strasbg.fr
(130.79.128.5) or via http://cdsweb.u-strasbg.fr/cgi-bin/qcat?J/A+A/}

}

\maketitle

\section{Introduction}

During the last decade considerable progress has been made in
modeling stellar atmospheres over large ranges of effective
temperatures, gravities and metallicities (see, e.g., \citet{H03a}
and references therein for the {\tt PHOENIX} models; \citet{CK03}
for {\tt ATLAS} models; \citet{P03} and \citet{G03} for {\tt
MARCS} models). While theoretical spectra show good agreement with
observations over a wide area in the HR diagram, late-type giants
are still thought to be one of the challenging exceptions
\citep[e.g.,][]{BCP98}.

The contribution from late-type giants on the Red Giant Branch
(RGB) and Asymptotic Giant Branch (AGB) is important in many
astrophysical contexts related to intermediate-age and old stellar
populations, thus  correct representation of their atmospheres and
spectra is of crucial importance. However, all state-of-art
stellar atmosphere models use a number of simplifications related
to the physics, and thus it is very important to know how well the
different theoretical models (e.g., {\tt PHOENIX}, {\tt MARCS},
{\tt ATLAS}) can reproduce photometric features of real stars, in
particular those for which reliable fundamental stellar parameters
(such as $T_{\rm eff}$, ${\rm log}\,g$, metallicity) are known.

The contribution of this study is three-fold. We provide a grid of
synthetic broad-band photometric colors for late-type giants,
based on the new {\tt PHOENIX} library of synthetic spectra
(Hauschildt et al. 2005, in preparation). The new {\tt PHOENIX}
library is an update and extension of the previous NextGen library
of synthetic spectra \citep{ng-hot, ng-giants} to lower effective
temperatures and metallicities. The major improvements of the
current models with respect to NextGen are updated equation of
state data and updated molecular opacities and line list, e.g.,
water and TiO lines (Sect.~\ref{models}). The new grid of
photometric colors covers $T_{\rm eff}=3000\dots5000$\,K, $\log
g=-0.5\dots{+3.5}$, and $\MoH=+0.5\dots{-4.0}$.

We also make a detailed investigation of the influence of model
parameters on the resulting broad-band photometric colors, namely,
the effects of molecular opacities, gravity, microturbulent
velocity, stellar mass, and the treatment of convection. This
analysis is done for colors at Solar metallicity and covers a wide
range of effective temperatures and gravities typical for
late-type giants ($T_{\rm eff}=3000\dots4800$\,K and $\log
g=0.0\dots{+3.0}$). To investigate the effects of convection on
the broad-band photometric colors, we calculate a full 3D
hydrodynamic model of a prototypical late-type giant ($T_{\rm
eff}\simeq 3700\,\mathrm{K}$, $\log g=1.0$, and ${\rm \MoH}=0.0$)
using the 3D model atmosphere code {\tt CO$^5$BOLD}, and provide a
comparison of 3D colors with those obtained using a standard 1D
model atmosphere.

Finally, we make an extensive comparison of the synthetic
broad-band photometric colors with observations of late-type
giants, and with empirical as well as theoretical $T_{\rm
eff}$--color and color--color relations available from the
literature. This comparison is done for colors at Solar
metallicity. For this purpose we employ a new {\tt PHOENIX} grid
of synthetic photometric colors, together with colors calculated
employing {\tt MARCS} (Plez 2003, private communication) and {\tt
ATLAS} \citep{CK03} model atmospheres. In order to compare
synthetic colors with observations, we derive a new $T_{\rm
eff}$--$\log g$--color relation employing published observations
of a homogeneous sample of late-type giants in the Solar
neighborhood, with effective temperatures available from
interferometry and surface gravities obtained using the $T_{\rm
eff}$--$\log g$ relation of \citet{H00}. We provide several new
semi-empirical $T_{\rm eff}$--$\log g$--color scales, which are
based on the $T_{\rm eff}$--$\log g$ relation of \citet{H00} and
synthetic colors from {\tt PHOENIX} (this work), {\tt MARCS}, and
{\tt ATLAS} model atmospheres. We also make a brief comparison of
the observed and synthetic spectra of late-type giants in order to
clarify what causes the differences between the observed and
synthetic photometric colors.

The paper is structured as follows. {\tt PHOENIX} stellar
atmosphere models, new synthetic spectra and broad-band
photometric colors are presented in Sect.~\ref{models}. A detailed
analysis of the effects of various model parameters on the
photometric colors is given in Sect.~\ref{effects}. We also
present here the first results of our exploratory 3D study of the
role of convection in late-type giants. A sample of nearby
late-type giants and astrophysical parameters of individual stars
are discussed in Sect.~\ref{sample}. Here we also analyze the role
of systematic effects related to the interferometric derivation of
angular diameters and effective temperatures. New empirical
$T_{\rm eff}$--$\log g$--color scales are derived in
Sect.~\ref{teffscales}, where we also provide a comparison of the
new synthetic colors with observations and $T_{\rm eff}$--color
and color--color relations from the literature. This section also
contains a comparison of observed and synthetic spectra calculated
with {\tt PHOENIX} and {\tt MARCS} model atmosphere codes.

This study deals with the photometric colors at solar metallicity;
the analysis of colors at sub-solar metallicities and effects of
metallicity are discussed in a companion paper \citep[][]{K04}.

\section{{\tt PHOENIX} models, spectra and synthetic colors of late-type giants\label{models}}

The {\tt PHOENIX} code is a very general non-LTE (NLTE) stellar
atmosphere modeling package \citep[]{phhs392, phhcas93,
phhnovfe295, faphh95, phhnov96, snefe296, parapap, parapap2, jcam,
LimDust} which can handle extremely complex atomic models as well
as line blanketing by hundreds of millions of atomic and molecular
lines. This code is designed to be both portable and flexible: it
is used to compute model atmospheres and synthetic spectra for,
e.g., novae, supernovae, M, L, and T dwarfs, irradiated
atmospheres of extrasolar giant planets, O to M giants, white
dwarfs and accretion disks in Active Galactic Nuclei (AGN). The
radiative transfer in {\tt PHOENIX} is solved in spherical
geometry and includes the effects of special relativity (including
advection and aberration) in the modeling.

\subsection{{\tt PHOENIX} stellar atmosphere models\label{models1}}

For our model calculations, we use the general-purpose stellar
atmosphere code {\tt PHOENIX} (version 13).  Details of the
numerical methods are given in the above references.

One of the most important recent improvements of cool stellar
atmosphere models is that new molecular line data  have become
available which have improved the fits to observed spectra
significantly. The combined molecular line database includes about
700 million lines. The lines are selected for every model from the
master line list at the beginning of each model iteration to
account for changes in the model structure.  Both atomic and
molecular lines are treated with a direct opacity sampling method
(dOS).  We do {\em not} use pre-computed opacity sampling tables,
but instead dynamically select the relevant LTE background lines
from master line lists at the beginning of each iteration for
every model and sum the contribution of every line within a search
window to compute the total line opacity at {\em arbitrary}
wavelength points. This approach also allows detailed and depth
dependent line profiles to be used during the iterations. This is
important in situations where line blanketing and broadening are
crucial for the model structure calculations and for the
computation of the synthetic spectra.

Although the direct line treatment seems at first glance
computationally prohibitive, it leads to more accurate models.
This is due to the fact that the line forming regions in cool
stars span a huge range in pressure and temperature so
that the line wings form in very different layers than the line
cores. Therefore, the physics of line formation is best modeled by
an approach that treats the variation of the line profile and the
level excitation as accurately as possible. To make this method
computationally more efficient, we employ modern numerical
techniques, e.g., vectorized and parallelized block algorithms
with high data locality \citep{parapap}, and use parallel
computers for the model calculations.

In the model grid used in this paper, we have included a constant
statistical velocity field, $\xi = 2$~km~s$^{-1}$, which is
treated like a microturbulence.  The choice of lines is dictated
by whether they are stronger than a threshold $\Gamma\equiv
\chi_l/\kappa_c=10^{-4}$, where $\chi_l$ is the extinction
coefficient of the line at the line center and $\kappa_c$ is the
local b-f absorption coefficient \cite[see][ for details of the
line selection process]{ng-hot}. This typically leads to about
10--250$\times 10^{6}$ lines which are selected from the  master
line lists.  The profiles of these lines are assumed to be
depth-dependent Voigt or Doppler profiles (for very weak lines).
Details of the computation of the damping constants and the line
profiles are given in \cite{vb10pap}. We have verified in test
calculations that the details of the line profiles and the
threshold $\Gamma$ do not have a significant effect on either the
model structure or the synthetic spectra.

The equation of state (EOS) is an updated version of
the EOS used in \citet{LimDust}. We include about 1000 species
(atoms, ions and molecules) in the EOS. The EOS calculations
themselves follow the method discussed in \citet{faphh95}.  For
effective temperatures, $T_{\rm eff} < 2500$ K, the formation of
dust particles has to be considered in the EOS. In our models we
allow for the formation (and dissolution) of a variety of grain
species. For details of the EOS and the opacity treatment see
\cite{LimDust}.

In this work we use a setup of the microphysics that gives the
currently best fits to observed spectra of M, L, and T dwarfs for
the low $T_{\rm eff}$ regime and that also updates the
microphysics used in the NextGen \cite{ng-hot,ng-giants} model
grid. The water lines are taken from the AMES calculations
\citep{ames-water-new}; this list gives the best overall fit to
the water bands over a wide temperature range. TiO lines are taken
from \cite{ames-tio} for similar reasons. The overall setup is
similar to the one described in more detail in \cite{LimDust}.

\begin{table}[tb]
\caption[]{Zero points of photometric color indices in the
Johnson-Cousins-Glass system (see Sect.~\ref{models3} for
details). References: 1.~\citet{BCP98}; 2.~\citet{CK94}; 3.~{\tt
PHOENIX} (NLTE), this work. \label{zeropoints}}
\setlength\tabcolsep{4.5pt}
\begin{tabular}{cccccccc}
\hline
\noalign{\smallskip}
$B\!-\!V$ & $V\!-\!R$ & $V\!-\!I$ & $V\!-\!K$ & $J\!-\!H$ & $J\!-\!K$ & $K\!-\!L$ & Ref    \\
\noalign{\smallskip}
\hline
\noalign{\smallskip}
   0.606   &   0.548   &   1.268   &   4.906   &   1.102   &   2.247   &   1.887   & 1    \\
   0.601   &   0.555   &   1.261   &   4.887   &   1.102   &   2.249   &   1.861   & 2   \\
\noalign{\smallskip}
\hline
\noalign{\smallskip}
   0.606   &   0.559   &   1.280   &   4.913   &   1.111   &   2.255   &   1.861   & 3    \\
\noalign{\smallskip}
\hline
\end{tabular}
\end{table}

\subsection{{\tt PHOENIX} grid of synthetic spectra for late-type giants\label{models21}}

The new grid of photometric colors (Sect.~\ref{models3}) is based
essentially on the new {\tt PHOENIX} library of synthetic spectra
(Hauschildt et al. 2005, in preparation\footnote{The spectra
are available at the following URL: ftp://\\
ftp.hs.uni-hamburg.de/pub/outgoing/phoenix/GAIA/v2.6.1/.}). To
summarize briefly, the spectra were calculated under the
assumption of spherical symmetry and LTE, with a typical spectral
resolution of $0.2$\,nm (which gradually degrades towards infrared
wavelengths). Microturbulent velocity was set to
$\xi=2$~km\,s$^{-1}$ (see discussion in Sect.~\ref{effects32}).
All models assumed spherical symmetry, therefore a mass of the
model star ($M_{\star}$) had to be specified: for all models in
the grid $M_{\star}=1\,M_{\odot}$ was used. Though the effect of
stellar mass on the broad-band colors is generally small, one
still has to be careful when using the models for stars with
significantly different masses (see Sect.~\ref{effects3}).

The models in this grid were calculated using a mixing length
parameter $\alpha_{\rm ML}\equiv l/H_{\rm p}=2.0$ ($l$ is the
mixing length and $H_{\rm p}$ is the local pressure scale height),
calibrated for M-type pre-main sequence objects and dwarf stars
\citep{L03}. This choice of mixing length may not be optimal for
giants, but it seems that changes in the emerging spectra due to
differences in the mixing length are minor within a framework of
1D model atmospheres. Note however, that in reality convection may
have a significant influence on the broad-band photometric colors,
because of convective overshoot into the outer atmospheric layers,
as it is hinted by our 3D modeling of a late-type giant
(Sect.~\ref{effects4}).

\begin{figure*}[tbp]
\centering
\includegraphics[width=18cm] {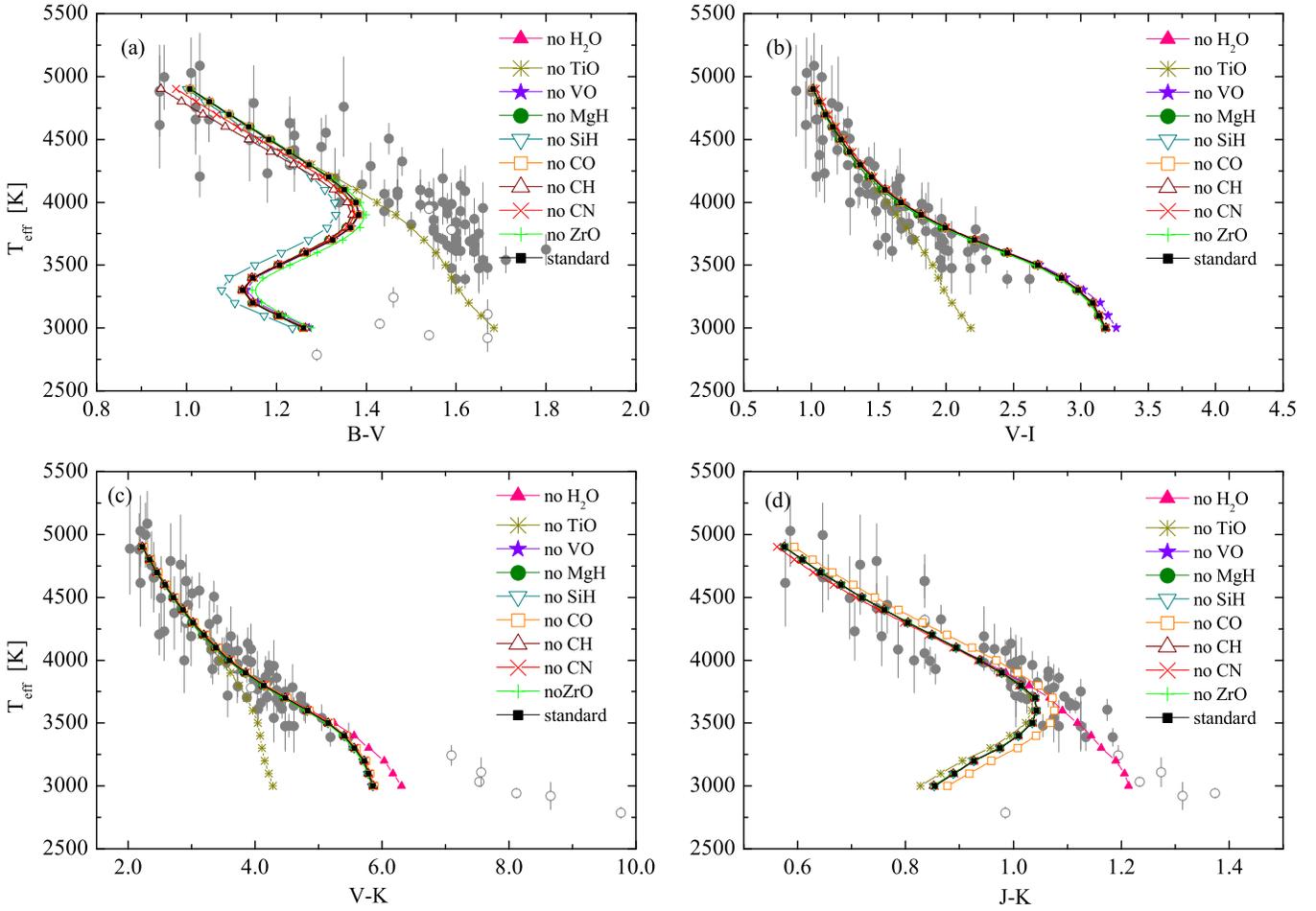}
\caption[]{Influence of various molecular bands on synthetic
photometric colors in different $T_{\rm eff}$--color planes (all
for $\log g=3.0$). Filled circles are late-type giants from
Table~\ref{samplestars}, open circles are stars from the sample of
\citet{P98} (see Sect.~\ref{sample} for details; stars are only
plotted to indicate the spread in the observed $T_{\rm
eff}$--color sequences, not for a detailed comparison). Thin lines
with symbols are synthetic {\tt PHOENIX} colors with certain
molecules excluded from the spectral synthesis calculations.
Symbols are spaced at every $100$\,K.\label{nomoleculeTC}}
\end{figure*}

\subsection{Synthetic {\tt PHOENIX} broad-band colors\label{models3}}

The broad-band colors were calculated from synthetic spectra in
the Johnson-Cousins-Glass system, using filter definitions from
\citet{B90} for the Johnson-Cousins \emph{BVRI} bands and
\citet{BB88} for the Johnson-Glass \emph{JHKL} bands. Conversion
of instrumental magnitudes to the standard Johnson-Cousins-Glass
system was done using zero points derived from the synthetic
colors of Vega (equating all color indices of Vega to zero). The
Vega spectrum used for this purpose was calculated with the {\tt
PHOENIX} code employing a full NLTE treatment. Adopted atmospheric
parameters were identical to those used by \citet{CK94}: $T_{\rm
eff}=9550$\,K, $\log g=3.95$, metallicity $\MoH=-0.5$ and
microturbulent velocity  $\xi=2\, \rm{km\,s^{-1}}$. Detailed Vega
abundances were taken from \citet{CK94}. Derived zero points of
color indices are given in Table~\ref{zeropoints}, together with
those from \citet{BCP98}. The latter were obtained using observed
and theoretical color indices of Vega and Sirius, with theoretical
colors computed employing the same filter transmission curves as
used in this work. For the purposes of comparison, we calculated
zero points using the Vega spectrum from \citet{CK94}, which are
also given in Table~\ref{zeropoints}. The agreement between the
three sets of zero points is generally very good. There is an
indication that the {\tt PHOENIX} zero points tend to be slightly
larger than those of \citet{BCP98}, though the differences are
typically $\la 0.01$~mag. The discrepancies are slightly larger
between the {\tt PHOENIX} zero points and those calculated from
the Vega spectrum of \citet{CK94}, though they are also $\la
0.02$~mag.

The final grid of synthetic photometric colors of late type giants
is given in Table~2 (available in electronic form
only), and covers the following parameter space: $T_{\rm
eff}=3000\dots5000$\,K (with a step of $\Delta T_{\rm eff} =
100$\,K), $\log g=-0.5\dots{+3.5}$ ($\Delta\log g = 0.5$) and
$\MoH=+0.5\dots-4.0$ ($\Delta\MoH=0.5$).

\subsection{{\tt MARCS} and {\tt ATLAS} spectra and colors}

For the comparison of synthetic colors with observations we also
used colors produced with {\tt MARCS} and {\tt ATLAS} model
atmospheres. {\tt MARCS} spectra employed in this work were kindly
provided to us by B. Plez (private communication, 2003). Models
were calculated in the plane-parallel geometry, using the mixing
length parameter $\alpha_{\rm ML}=1.5$ (for more details about the
{\tt MARCS} models see Plez 2003). Broad-band photometric colors
were calculated using the procedure described in
Sect.~\ref{models3}.

Synthetic {\tt ATLAS} broad-band photometric colors were taken
from \citet{CK03}. Models were calculated under the assumption of
plane-parallel geometry, using the turbulent velocity $\xi=2\,
\rm{km\,s^{-1}}$ and mixing length parameter $\alpha_{\rm
ML}=1.25$. Water and TiO opacities used were identical to those
employed by us in the calculation of the {\tt PHOENIX} grid (see
Sect.~\ref{models1}).

\section{The influence of astrophysical processes and model parameters
         on synthetic photometric colors\label{effects}}

The thermal structure of a model atmosphere is governed a by a
number of input parameters (such as stellar mass, gravity,
metallicity, etc.). Indeed, all of them have a direct influence on
the emerging spectrum and photometric colors. In this Section we
investigate the role and possible extent of such effects, to
provide a theoretical grounding for the comparisons of model
predictions with observations (Sect.~\ref{teffscales}).

\subsection{Effects of molecular opacities}

\begin{figure}[t]
\centering
\includegraphics[width=8.8cm] {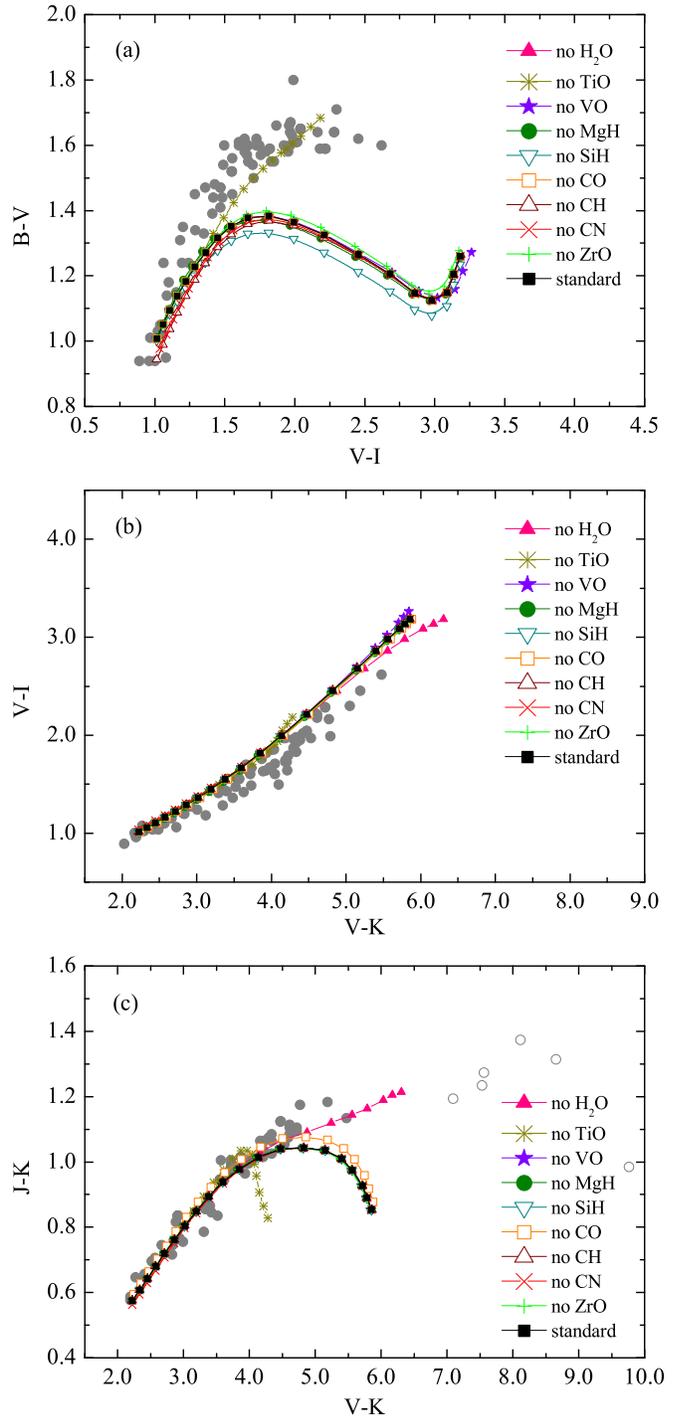}
\caption[]{Same as Fig.~\ref{nomoleculeTC} but in the color--color
planes.\label{nomoleculeCC}}
\end{figure}

Spectra of the late-type stars are heavily blended by various
molecular lines and bands, especially at the effective
temperatures below $T_{\rm eff}\sim4000$\,K (${\rm H}_{2}{\rm O}$,
TiO, VO, CO, etc.). To investigate the extent of these effects we
produced a number of synthetic spectra with certain molecules
`switched off' during the spectral synthesis calculations. In this
procedure model structures were calculated as usual, i.e.,
employing opacities of all relevant molecules, as in the
calculation of all standard spectra in the model grid discussed in
Sect.~\ref{models}. The spectral synthesis, however, was done
subsequently for several different cases without using opacities
of certain key molecules (${\rm H}_{2}{\rm O}$, TiO, VO, MgH, SiH,
CO, CH, CN, and ZrO). The resulting spectra are thus different
from the standard ones, as lines/bands of certain molecules are
not seen in the spectra, while the physics involved in the
calculations of the model structures is identical in all cases.
The differences between the synthetic colors calculated from these
spectra and the standard spectrum (i.e., the one with all
opacities `on') clearly show the effect of a particular molecule
on a given photometric color (Fig.~\ref{nomoleculeTC}, $T_{\rm
eff}$--color planes; Fig.~\ref{nomoleculeCC}, color--color
planes).

Obviously, TiO is by far the most influential molecule in the
optical wavelength range. It is responsible for the turn-off
toward the bluer color in the $T_{\rm eff}$--$(B-V)$ plane at
$T_{\rm eff}\sim 3700$\,K, and for the significant reddening of
photometric colors below $T_{\rm eff}\sim 4000$\,K in the $T_{\rm
eff}$--$(V-I)$ and $T_{\rm eff}$--$(V-K)$ planes. The reddening of
$V-I$ and $V-K$ colors is caused by the increasing strength of TiO
lines in the $V$ band with decreasing $T_{\rm eff}$: since the $I$
band is much less influenced by TiO bands than the $V$ band (while
the $K$ is not affected), $V-I$ and $V-K$ colors gradually become
redder at low effective temperatures (the reason why $B-V$ gets
bluer below $T_{\rm eff}\sim3700$\,K is explained in
Sect.~\ref{models2}). There is also some influence of TiO on
$J-K$, due to TiO lines in the $J$ band.

Water opacity is significant in all near-infrared photometric
bands. Most affected is the $J-K$ color, as ${\rm H}_{2}{\rm O}$
lines are strong both in $J$ and $K$ bands. Note however that
effects of ${\rm H}_{2}{\rm O}$ become noticeable only at lower
effective temperatures ($T_{\rm eff}\leq3700$\,K). There is a weak
influence of CO and CN on the $J-K$ color too, due to CO lines on
the edges of $J$ and $K$ bands, and CN in the $J$ band (the latter
comes into play only at $T_{\rm eff}\ga4000$\,K).

\begin{figure*}[tb]
\centering
\includegraphics[width=18cm] {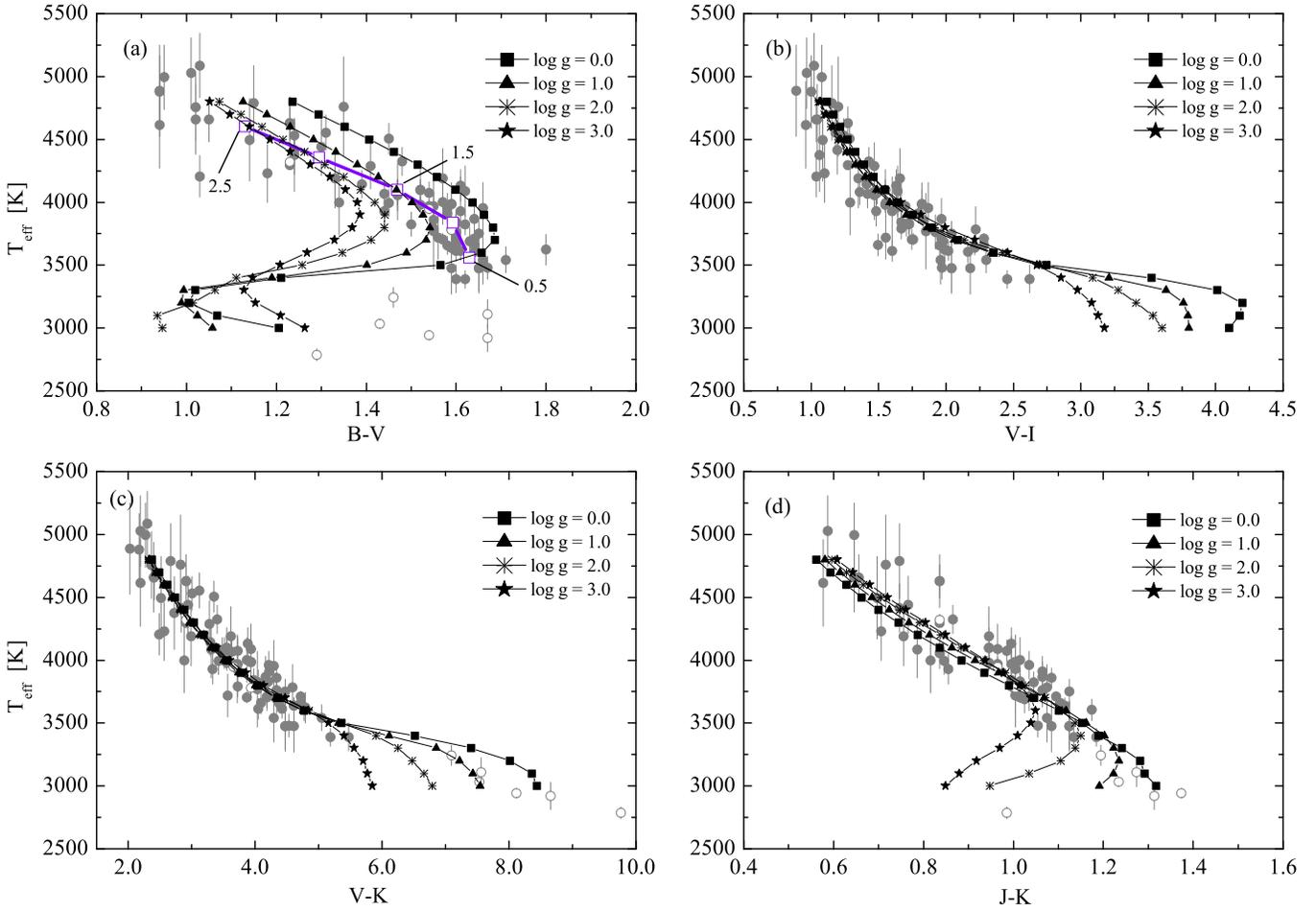}
\caption[]{Influence of gravity on broad-band photometric colors
in the $T_{\rm eff}$--color planes. Filled circles are late-type
giants from Table~\ref{samplestars}, open circles are stars from
the sample of \citet{P98} (see Sect.~\ref{sample} for details;
stars are only plotted to indicate the spread in the observed
$T_{\rm eff}$--color sequences, not for a detailed comparison).
Thin lines with symbols are the new {\tt PHOENIX} colors at
different $\log g$ values (symbols are spaced at every $100$\,K).
Thick solid line with open rectangles in panel (a) shows $T_{\rm
eff}$--$\log g$--$(B-V)$ locii for the sample of late-type giants
in the solar neighborhood (Sect.~\ref{sample}), with gravities
assigned according to the $T_{\rm eff}$--$\log g$ relation of
\citet{H00}. Numbers next to the open rectangles are $\log g$
values, with symbols plotted at every $0.5$\,dex in $\log g$. Note
a strong sensitivity of the $B-V$ color on $\log g$ in the entire
effective temperature range. \label{gravityTC}}
\end{figure*}

Optical photometric colors are also noticeably affected by SiH,
MgH, VO, CH, CN, and ZrO. The strongest lines of SiH and MgH are
located in the blue part of the spectrum ($\sim$400--450\,nm and
$\sim$450--550\,nm, respectively), affecting the $B$ band flux in
the former case and $V$ band flux in the latter. The $B-V$ color
is rather strongly influenced also by CH and to a smaller extent
by CN, due to CH lines in the $B$ band (especially the G-band at
$\sim430$\,nm), and CN bands both in $B$ and $V$. In both cases
the effect sets in at higher temperatures ($T_{\rm
eff}\ga3800$\,K). There is a weak influence of ZrO on the $B-V$
color too, due to ZrO bands at $\sim$550--700\,nm. The strongest
VO lines are in the wavelength range of $\sim$700--900\,nm, thus
mostly influencing the $I$ band.

The trends in the color--color diagrams essentially follow those
in the $T_{\rm eff}$--color planes (Fig.~\ref{nomoleculeCC}). It
is interesting to note that the $(V-I)$--$(V-K)$ plane is
relatively little influenced by the molecular opacities (relative
to the observed spread in photometric colors), even those of TiO
(as both $V-I$ and $V-K$ colors get considerably bluer without
TiO, the overall trend is little affected).

In general, the influence of different molecules on the broad-band
photometric colors is small at $T_{\rm eff}\geq4000$\,K, except
for some influence of SiH, CH ($B$ band), CN ($B$, $V$), and CO
($J$, $K$).

\subsection{Effects of gravity\label{models2}}

\begin{figure}[tbp]
\centering
\includegraphics[width=8.8cm] {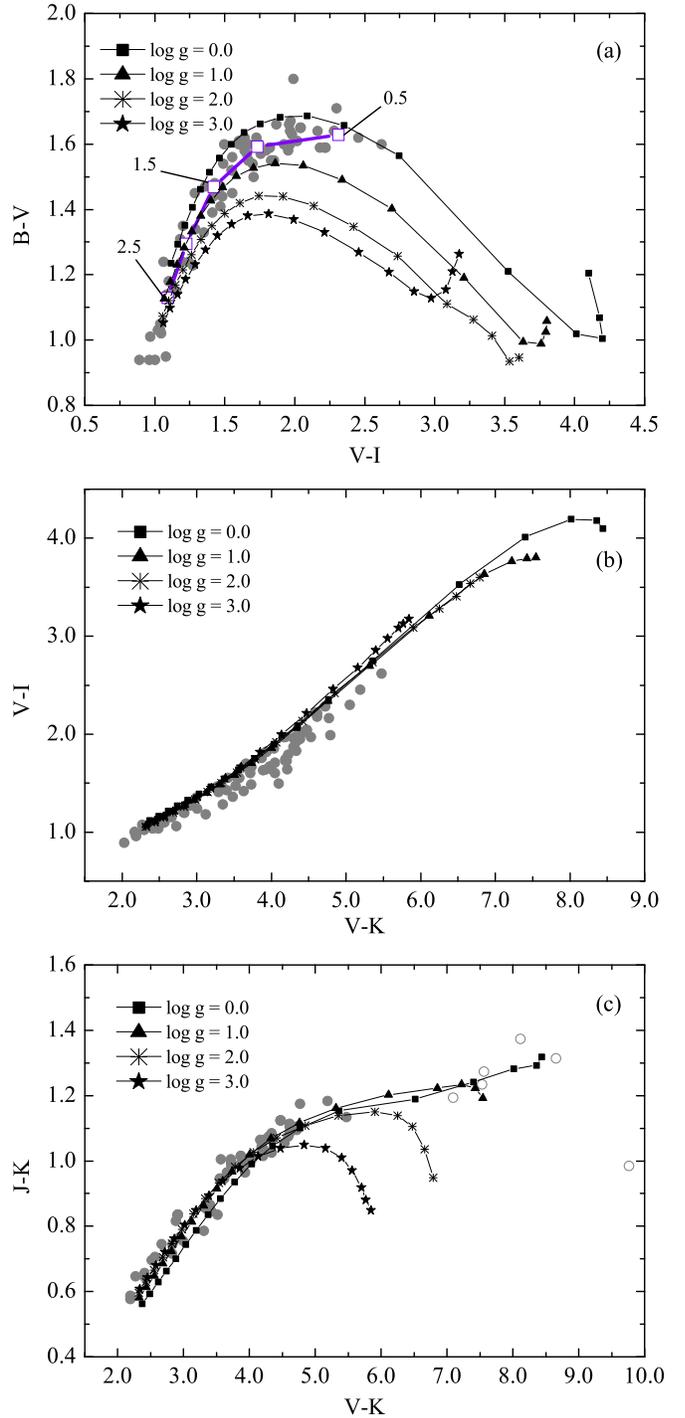}
\caption[]{Same as Fig.~\ref{gravityTC} but in the color--color
planes.\label{gravityCC}}
\end{figure}

For a given stellar mass and effective temperature, surface
gravity defines the stellar radius and thus the extension and
structure of the outer photosphere, where an important fraction of
the emerging spectral flux is formed. Indeed, the effects of
gravity must be significant in late-type giants, especially at low
$T_{\rm eff}$. The actual extent of these effects on broad-band
photometric colors can be seen in Figs.~\ref{gravityTC} and
\ref{gravityCC}, which show the behavior of photometric colors in
various $T_{\rm eff}$--color (Fig.~\ref{gravityTC}) and
color--color planes (Fig.~\ref{gravityCC}) at different gravities.

The influence of gravity on broad-band photometric colors is
generally small above $\sim$3700\,K. $V-K$ is especially robust in
this sense; little sensitivity is also seen for $V-I$ and $J-K$,
both in the $T_{\rm eff}$--color and color--color domains. At
these relatively high temperatures, only a few molecules, e.g.,
H$_2$, CO, CH and SiH, survive at gravities between $0.0$ and
$3.0$, thus having little effect on the colors. At lower
temperatures, $\la 3500$\,K, the gravity has a much larger effect
on the chemistry of the atmosphere; for example, water vapor
begins to form for $\log g=3.0$, whereas it is absent at $\log
g=0.0$. Therefore, the colors are more gravity dependent at
$T_{\rm eff}\sim3500$\,K than at $\sim$3700\,K.

Obviously, the influence of gravity is strongest for the $B-V$
color, which is clearly reflected in the $T_{\rm eff}$--$(B-V)$
and $(B-V)$--$(V-I)$ planes
(Figs.~\ref{gravityTC}a,\,\ref{gravityCC}a). This results from the
fact that the temperature structure of the atmospheres changes
with the gravity: at low gravity, the atmospheres are more
extended giving lower gas temperatures than at higher gravities.
For example, at $T_{\rm eff}\sim3500$\,K the outermost temperature
of the atmosphere models is about $2100$\,K for $\log g=0.0$
whereas it is $\sim$2300\,K for $\log g=3.0$. The differences in
the model stratifications have a strong influence on molecule
formation, which is much more efficient in the cooler, low gravity
models. Since the $B-V$ color is very sensitive to molecular
opacities (TiO, SiH, etc.), the influence of gravity on this color
is strong.

Another feature clearly seen in the $T_{\rm eff}$--$(B-V)$ and
$(B-V)$--$(V-I)$ diagrams at all gravities is a `turn-off' towards
the bluer colors in $B-V$ at $\sim$3700--3900\,K. This inversion
occurs because the $V$-band flux is strongly affected by the TiO
opacity, which is growing rapidly with decreasing $T_{\rm eff}$.
The TiO lines are much weaker in the $B$-band, thus the decrease
of $B$-band flux is essentially governed by the shift of the
maximum of the emitted spectral energy towards longer wavelengths
with decreasing $T_{\rm eff}$. The net effect is that below
$T_{\rm eff}\sim4200$\,K the total flux in the $V$-band decreases
faster with decreasing $T_{\rm eff}$ than in the $B$-band, causing
the turn-off in $B-V$ at $T_{\rm eff}\sim3700$--$3900$\,K.

Interestingly, this effect has been noticed observationally nearly
four decades ago \citep[e.g.,][]{J66,W67}. \citet{W67} has found
that observed $B-V$ color of late-type giants reaches its maximum
value of $B-V\simeq1.65$ at around M2~III in $T$--$(B-V)$ plane
(here $T$ is a black-body temperature measured from the black-body
fit to two pseudo-continuum points in the 0.75--1.04 micron
range). This corresponds to $T_{\rm eff}\simeq3710$\,K according
to the effective temperature -- spectral type scale of
\citet{Pi98}. At lower temperatures, the observed $B-V$ colors
stay essentially unchanged for M0~III--M4~III, then become bluer
for later spectral types and finally turn to the red again for the
coolest giants. It should be noted that below $T_{\rm
eff}\sim3500$\,K the bluest observed $B-V$ color in the sample of
\citet{W67} corresponds to $B-V\sim1.4$, while theoretical models
predict $B-V\la1.2$. Unfortunately, the lack of knowledge of
effective temperatures (or precise spectral types) for the
late-type giants in the sample of \citet{W67} does not allow us to
make a direct comparison of his findings with the synthetic colors
calculated using current stellar atmosphere models in the $T_{\rm
eff}$--$(B-V)$ plane.

\begin{figure*}[tbp]
\centering
\includegraphics[width=18cm] {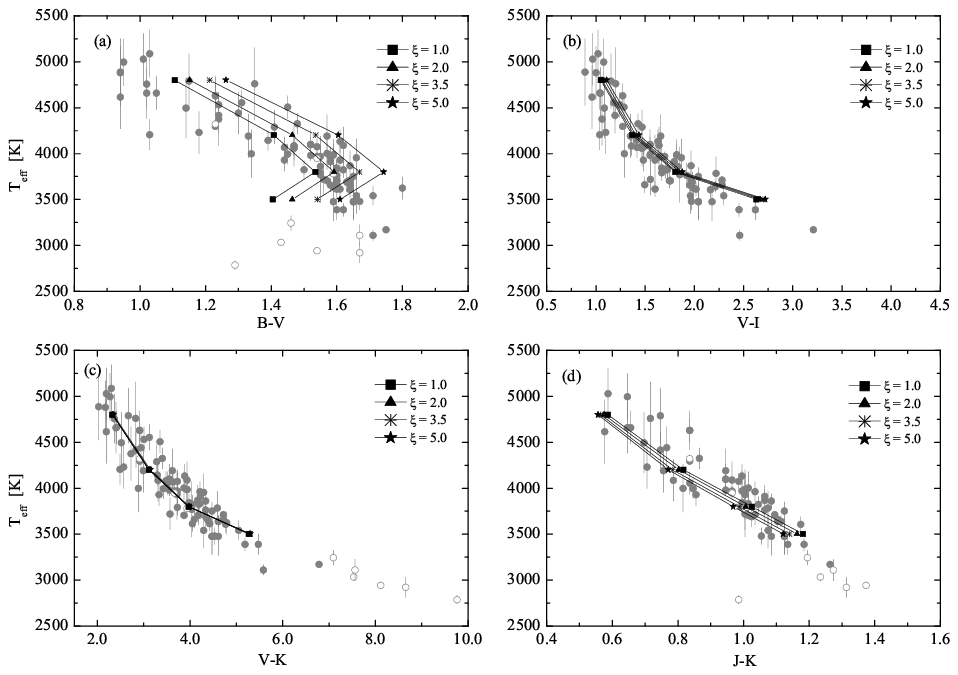}
\caption[]{Influence of microturbulent velocity on synthetic
photometric colors in different $T_{\rm eff}$--color planes (all
for $\log g=0.5$). Filled circles are late-type giants from
Table~\ref{samplestars}, open circles are stars from the sample of
\citet{P98} (see Sect.~\ref{sample} for details; stars are only
plotted to indicate the spread in the observed $T_{\rm
eff}$--color sequences, not for a detailed comparison). Thin lines
with symbols are synthetic {\tt PHOENIX} colors at different
microturbulent velocities. \label{xiTC}}
\end{figure*}

A similar effect can be seen in the mean observed $B-V$ colors of
late-type giants provided by \citet{J66}, with the turn-off at
around M3~III ($T_{\rm eff}\simeq3630$\,K on the scale of Pickles
1998), and with the maximum value of $B-V\simeq1.60$. Observed
colors of late-type giants in the sample of \citet{P98} show hints
of this turn-off too (Fig.~\ref{gravityTC}). However, in all these
cases the effect seems to set in at somewhat lower effective
temperatures than predicted by the theoretical models: typically,
the turn-off in theoretical colors occurs at $T_{\rm
eff}\sim3700-3800$\,K, while the observations point to $T_{\rm
eff}\sim3600-3700$\,K (see Sect.~\ref{teffscales} below for a
detailed discussion). Note however, that stars in the sample of
\citet{P98} are all variable and this may indeed influence their
$B-V$ colors. This may be the case with the coolest giants in the
sample of \citet{W67} too. Reconstruction of photometric colors of
a `parent star' (i.e., a static star with the same parameters as
the variable) is not trivial in case of long-period variables
\citep{HSW98}, since a simple averaging of the photometric colors
over the pulsational cycle is generally not appropriate. This may
thus easily bias the $T_{\rm eff}$--color scales at the lower
effective temperature end, i.e., below $T_{\rm eff}\la3500$\,K. It
is perhaps worthwhile to note in this respect that in our search
for the published interferometric effective temperatures of
late-type giants in the solar neighborhood we found no
non-variable giants with effective temperatures lower than $T_{\rm
eff}\sim3400$\,K.

It should be stressed that all colors are significantly influenced
by gravity below $T_{\rm eff}\sim3600$\,K, due to the onset of
rapid molecule formation at these temperatures. The `turn-off'
towards bluer colors seen in all color domains at $T_{\rm eff} \la
3500$\,K is thus produced by the increasing opacities of various
molecules (TiO, ${\rm H}_{2}{\rm O}$, VO, etc.). The position of
the `turn-off' is gravity dependent, since the models with lower
gravity are cooler leading to a higher rate of molecule formation.

Note however, that at these very low effective temperatures and
gravities the atmospheres of late-type giants become very
extended, thus stellar atmosphere models employing plane-parallel
geometry (e.g., {\tt ATLAS}) will be not adequate. In fact,
spherical models may be insufficient too, since non-spherical and
non-stationary phenomena (convection, variability, shock-waves,
mass loss, etc.) will become increasingly important in this
effective temperature and gravity domain (as hinted, for example,
by 3D models of the red supergiant Betelgeuse, Freytag et al.
2002).

\subsection{Effects of microturbulent velocity\label{effects32}}

\begin{figure}[tbp]
\centering
\includegraphics[width=8.8cm] {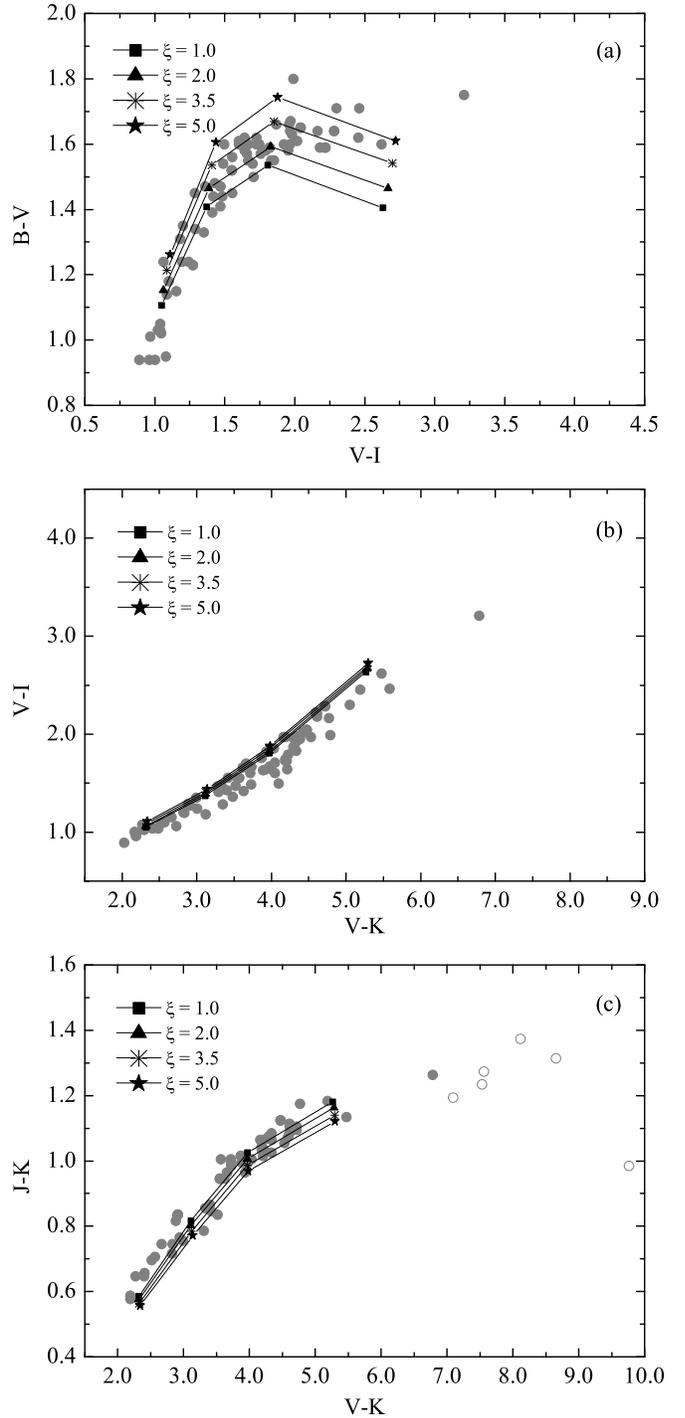}
\caption[]{Same as Fig.~\ref{xiTC} but in the color--color
planes.\label{xiCC}}
\end{figure}

\begin{figure}[tbp]
\includegraphics[width=8.8cm] {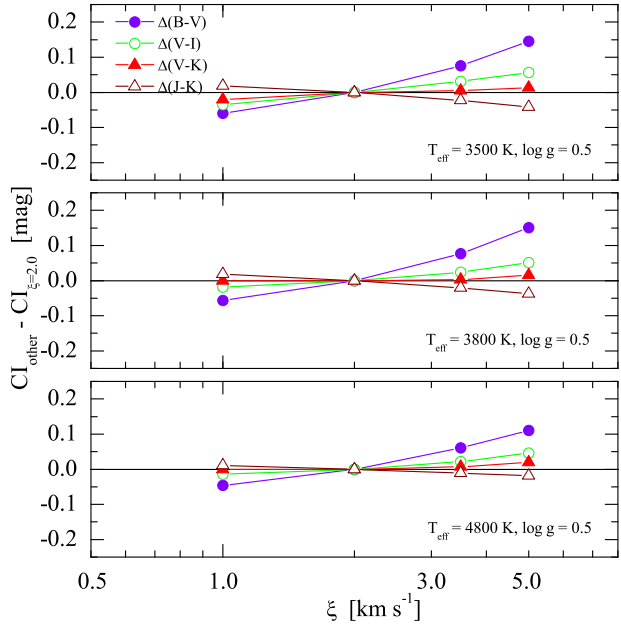}
\caption[]{Effect of microturbulent velocity ($\xi$) on broad-band
photometric colors at different effective temperatures (the
ordinate is the difference between the photometric color index for
a given microturbulent velocity and that for $\xi=2.0$\,km\,${\rm
s^{-1}}$). \label{xi} }
\end{figure}

Grids of synthetic colors produced with {\tt PHOENIX}, {\tt
MARCS}, and {\tt ATLAS} model atmospheres (Sect.~\ref{models3})
are based on the synthetic spectra calculated using a single value
of microturbulent velocity, $\xi=2.0$~km\,s$^{-1}$. Indeed,
variations around this value must be anticipated in real stars. To
investigate the effect of these variations on the broad-band
photometric colors we have calculated a set of {\tt PHOENIX}
models and spectra at several additional values of microturbulent
velocity, $\xi=1.0, 3.5, 5.0$~km\,s$^{-1}$. The results are
summarized in Figs.~\ref{xiTC}-\ref{xiCC}, which show the
influence of $\xi$ in the $T_{\rm eff}$--color and color--color
planes. Fig.~\ref{xi} provides a more detailed view on the
differences between colors calculated at any particular value of
microturbulent velocity and those at $\xi=2.0$~km\,s$^{-1}$, at
several effective temperatures.

Clearly, $B-V$ color is most sensitive to changes in
microturbulent velocity (Fig.~\ref{xiTC}a). The differences are
indeed significant, $\Delta(B-V)\sim0.2$\,mag at $T_{\rm
eff}=3600$\,K for colors corresponding to $\xi=1.0$ and
$5.0$~km\,s$^{-1}$. The flux in the $B$ and $V$ band is affected
by changes in $\xi$ (though the effect is considerably stronger in
the $B$ band), and in both cases the flux is lower at higher
microturbulent velocities, mostly due to the increasing line
blending with higher $\xi$. The effect is smaller but still
non-negligible in case of $V-I$ and $J-K$ (up to $\sim0.1$\,mag,
Figs.~\ref{xiTC}b-c). The only color that is essentially
unaffected by the changes in microturbulent velocity is $V-K$
(Fig.~\ref{xiTC}d). In this case the flux in $V$ and $K$ bands is
lower at higher values of $\xi$ by a comparable amount (mostly due
to increasing line widths of numerous atomic lines in the former
case and due to changes in the width of CO bands in the latter),
thus the $V-K$ color remains essentially unaffected.

In general, differences between colors corresponding to different
$\xi$ are largest at lower effective temperatures but they remain
significant even at relatively high temperatures, e.g., $T_{\rm
eff}\ga4500$\,K. The effect of microturbulent velocity is somewhat
smaller at higher gravities, which is related to the fact that at
higher $\log g$ spectral lines are broader than at lower gravities
and thus the relative broadening due to the increasing
microturbulent velocity (or vice versa) is smaller. Hence, the
emitted flux and thus the photometric colors are affected less
than at lower gravities.

Trends in the color--color diagrams follow those seen in the
$T_{\rm eff}$--color planes. The effect is largest in the
$(B-V)$--$(V-I)$ plane, especially at lowest effective
temperatures, due to the sensitivity of the $B-V$ color to changes
in the microturbulent velocity. The effect is also significant in
the $(J-K)$--$(V-K)$ plane where differences in the $J-K$ color
are comparable with the spread in the observed giant sequence. The
$(V-I)$--$(V-K)$ plane is little affected since the influence of
$\xi$ is small both in the case of the $(V-I)$ and $(V-K)$ color.

\subsection{Effects of stellar mass\label{effects3}}

\begin{figure}[tbp]
\includegraphics[width=8.8cm] {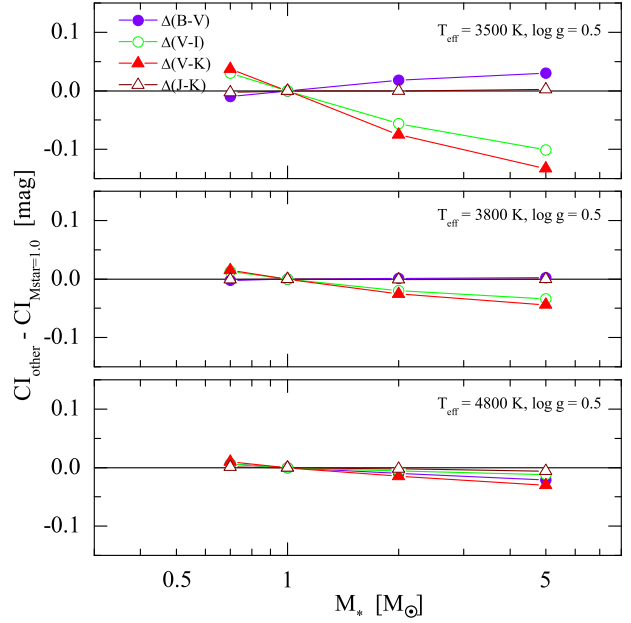}
\caption[]{Effect of stellar mass on broad-band photometric colors
at different effective temperatures (the ordinate is the
difference between the photometric color index for a given stellar
mass and that for $M_{\star}=1.0$\,${\rm M}_{\odot}$).
\label{mstar} }
\end{figure}

The influence of stellar mass on broad-band colors is illustrated
in Fig.~\ref{mstar}, where we plot differences between the
photometric colors corresponding to stars of different masses
(${\rm CI(M_{\star})} - {\rm CI(M_{\star}=M_{\odot})}$) at several
effective temperatures, for $\log g=0.5$ (the differences are
smaller at higher gravities). The effect of stellar mass is most
pronounced at lower effective temperatures ($T_{\rm eff} \la
4500$\,K), where differences in e.g. $V-K$ may reach
$\sim$0.1\,mag at $T_{\rm eff}\sim3500$\,K when comparing stars
with $M_{\star}=1.0$ and $5.0\,M_{\odot}$. The effect becomes
smaller at higher effective temperatures, but is for some colors
non-negligible even at $T_{\rm eff}\sim4800$\,K
(Fig.~\ref{mstar}).

It should be noted, however, that these differences do not simply
mimic the effect of gravity. It has been shown above that, for
instance, $B-V$ is rather sensitive to gravity, while $V-K$
remains essentially unaffected. The situation is opposite in terms
of sensitivity to stellar mass (Fig.~\ref{mstar}), which shows
that stellar interiors respond in a rather different way to
changes in mass and gravity. To a large extent this is determined
by the fact that the atmospheric structure essentially remains
unaltered with changing stellar mass (on the scale of optical
depth), though the outer layers are marginally hotter for
$M_{\star} = 5M_{\odot}$ than $1M_{\odot}$ (especially at lower
effective temperatures). Together with changes in the molecular
dissociation equilibria, this produces a slight shift of the
photometric colors towards the higher effective temperatures seen
in Fig.~\ref{mstar}. Since {\tt PHOENIX} spectra and colors are
calculated for the stellar mass of $M_{\star} = 1M_{\odot}$, these
effects should be taken into account when using them with stars of
considerably different masses.

\subsection{Influence of the treatment of convection\label{effects4}}

The new grid of {\tt PHOENIX} spectra/colors presented in this
work employs a mixing length parameter $\alpha_{\rm ML}=2.0$ which
was motivated by results from the 3D hydrodynamic modeling of
M-type pre-main sequence objects and dwarf stars \citep{L03}. All
1D standard model atmospheres of evolved giants discussed so far
in this paper predict that the structure of the optically thin
layers hardly depends on the assumed mixing length parameter: we
find that relative differences in the spectral flux for the {\tt
PHOENIX} models calculated with the mixing lengths of $\alpha_{\rm
ML}=1.5$ and $2.0$ are small (typically of the order of few tenths
of a percent). Differences in the broad-band colors are very small
too, typically within a few milimagnitudes. The situation with
{\tt MARCS} spectra and colors is quantitatively very similar (B.
Plez, private communication).

The simple reason for the insensitivity is that in the framework
of mixing length theory the convective zone is confined to
optically thick regions. However, the geometric distance between
the upper boundary of the formally convectively unstable region
(according to the Schwarzschild criterion) and optical depth unity
is not large. One might speculate that in a real star convection
may overshoot into the optically thin layers and thus may
influence the emergent spectrum.

\begin{figure}[tb]
\includegraphics[width=8.8cm]{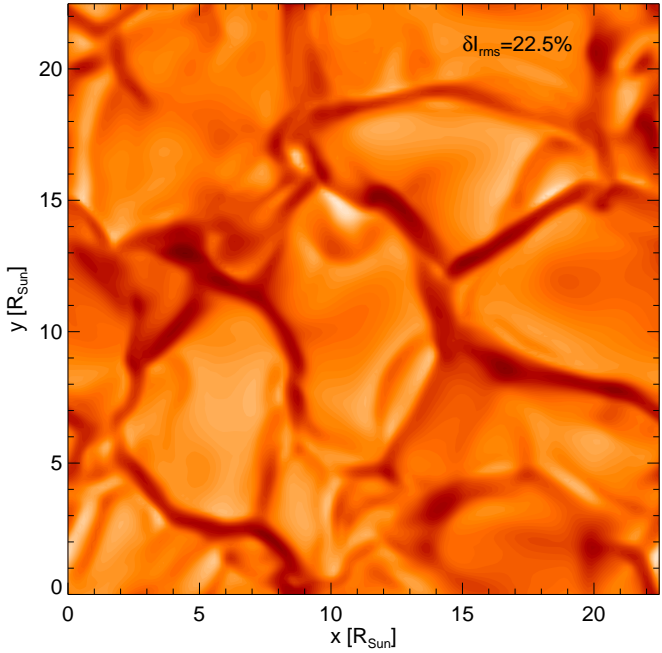}
\caption[]{Snapshot of the emergent white light intensity during
the temporal evolution of a hydrodynamical red giant model. Note
the presence of the granulation pattern, and its spatial scales.
The relative root-mean-squares contrast of the intensity amounts
to 22.5\,\% at this particular instant in
time.\label{f:giantgranulation}}
\end{figure}

To test the idea we performed an exploratory study, and
constructed a 3-dimensional hydrodynamical model atmosphere for a
prototypical late-type giant with $T_{\rm eff}\simeq 3700$\,K,
$\log g=1.0$, and ${\rm \MoH}=0.0$. For this purpose we employed
the radiation-hydrodynamics code {\tt CO$^5$BOLD} mainly developed
by B.~Freytag and M.~Steffen; for a description of the code see
\citet{WFSLH04}.  The model is a so-called `local box' model
employing grey radiative transfer and Cartesian geometry. The
computational box contains the optically thin atmosphere and the
upper part of the optically thick convective stellar envelope.
Details of the model will be discussed elsewhere (Ludwig et al.
2005, in preparation), while in this study we concentrate on
issues related to spectral properties.
Figure~\ref{f:giantgranulation} shows a snapshot of the emergent
intensity during the temporal evolution of the model. One
immediately realizes the presence of a granulation pattern.
Similar to 1D hydrostatic model atmospheres based on mixing length
theory the average vertical structure of the hydrodynamical model
is such that the stratification becomes formally convectively
stable already in optically thick layers. Hence, the granular
pattern is a consequence of intense convective overshooting. The
time-averaged relative root-mean-squares intensity contrast
amounts to $22.5$\,\%, which is larger than the contrast of solar
granulation ($18$\,\% white light contrast).

To obtain an estimate of the amount of the color changes due to
convection-related effects we performed spectral synthesis
calculations for the hydrodynamical model atmosphere, and compared
them to corresponding results from a standard 1D hydrostatic model
atmosphere based on mixing length theory but otherwise identical
input physics. In particular, we used the same equation of state
and grey {\tt PHOENIX} opacities in the hydrodynamical and
hydrostatic model. Turbulent pressure was neglected when solving
the hydrostatic equation.  In the hydrodynamical model turbulent
pressure makes a substantial contribution to the dynamical
balance. It leads to a lifting of the optically thin layers to
larger radii, and also alters to some extent their
pressure-temperature relation. The neglect of turbulent pressure
in the 1D model has nevertheless no consequences for our
comparison.  Due to the confinement of convection to the optically
thick layers no reasonable choice of the treatment of turbulent
pressure in the 1D model alters the pressure-temperature relation
of the optically thin layers, or might allow to emulate the
behavior of the hydrodynamical model.

\begin{figure}[tb]
\includegraphics[width=8.8cm]{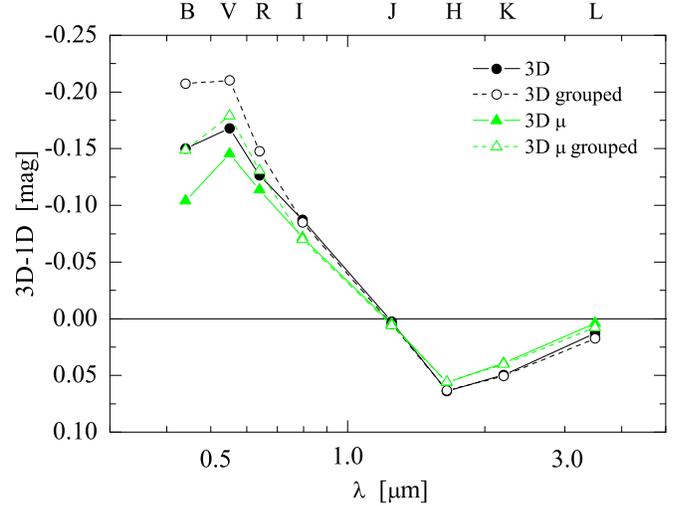}
\caption[]{Influence of surface granulation on the broad-band
photometric colors of the prototypical red giant, as reflected by
magnitude differences in various band-passes (indicated on the top
of the panel) between the predictions of 3D hydrodynamical and
classical 1D model atmospheres. See text for details.
\label{f:color3D}}
\end{figure}

Because of time stepping constraints related to short radiative
time scales the calculation of the hydrodynamical giant model was
very time consuming. We were only able to gather data of a rather
short time sequence limiting the obtainable statistical accuracy.
The hydrodynamical spatio-temporal model has
$150\times150\times100$ grid points ($X\times Y\times Z$, where
$Z$ is vertical), which corresponds to
$15750\times15750\times8602$\,${\rm Mm}^3$ geometrical box size,
and provides a total of $2.25\times10^6$ pressure-temperature
profiles. Together with the typical number of $\sim5\times 10^5$
wavelength points usually used in the framework of {\tt PHOENIX}
it renders the spectral synthesis problem intractable if one tries
to calculate spectra for all thermal profiles individually, and
subsequently averages them to obtain the observable spectrum.
Instead, we used an approximate approach and sorted the individual
thermal profiles into nine groups of increasing emergent white
light intensity. This corresponds in Fig.~\ref{f:giantgranulation}
to a sorting of points depending on whether they belong to an
inter-granular lane or increasingly brighter parts of a granule.
This classification turns out to group together thermally as well
kinematically (in terms of velocity) similar vertical
stratifications. The stratifications associated with a particular
group were then averaged on surfaces of equal optical depth. The
resulting nine stratifications were treated as standard 1D model
atmospheres for which spectral synthesis calculations were
performed.  The nine radiation fields were added weighted by their
respective surface area fraction. For the further comparison we
also obtained a `global average', by averaging over {\it all\/}
pressure-temperature profiles.

Figure~\ref{f:color3D} shows resulting flux ratios (expressed as
magnitude differences) between the 3D hydrodynamical and 1D
hydrostatic models integrated over various band-passes. We
corrected small mismatches in total flux of the individual
synthetic spectra by scaling them with a wavelength independent
factor to obtain the same nominal flux. The representation of
fluxes in Fig.~\ref{f:color3D} also allows to calculate color
differences 3D$-$1D by simply subtracting the magnitudes in
corresponding filters.  The global average (`3D') contains effects
on the radiation field merely associated with the different
vertical structure of the hydrodynamical atmosphere, while the
superposition of the radiation fields of the individual 9 groups
contains additional effects related to thermal inhomogeneities in
horizontal directions (`3D grouped'). Consequently, differences
3D$-$1D become somewhat larger in the `grouped' case. They are
most pronounced towards shorter wavelength due to the more
strongly non-linear dependence of the source function on
temperature so that horizontal temperature inhomogeneities can
more easily leave a noticeable imprint in the average emergent
radiation field.

Obviously, the grouping procedure described before ignores the
geometry of the flow for off-center positions on the stellar disk.
Each horizontal position is represented by a plane-parallel
infinitely extended model atmosphere, ignoring the neighborhood of
hot and cool areas. To include more accurately the 3D effects of
the center-to-limb variation we extended our classification scheme
by keeping the groups but calculated the average structures and
radiation fields for inclined viewing angles (three in total)
separately, including the full information about the flow
geometry.  Figure~\ref{f:color3D} shows that the 3D limb effects
tend to decrease the deviations 3D$-$1D. The lines `3D' and `3D
$\mu$' (for the case of including 3D limb effects) depict the case
where horizontal fluctuations are not explicitly accounted for.
Lines-of-sight at inclined viewing angles pass through parts of
the atmosphere belonging to different groups, thus leading to some
`mixing' between the groups which should reduce the effects of
horizontal fluctuations.  Indeed, the inclusion of horizontal
fluctuations lead to slightly more pronounced differences as
depicted by the lines `3D grouped' (3D limb effects ignored but
horizontal fluctuations included) and `3D $\mu$ grouped' (3D limb
effects and horizontal fluctuations included). However, all
approaches provide a rather similar picture indicating that the
difference in the mean vertical structure of 3D and 1D model is
the dominant factor for the differences in the band-integrated
fluxes (resp. colors).

The flux differences 3D$-$1D may lead to noticeable differences in
effective temperatures when they are derived from photometric
colors based on 1D standard model atmospheres (e.g., the
difference in $(V-K)$ of $\sim$0.2 translates to an offset in
$T_{\rm eff}$ by $\sim70$\,K). However, the changes in $T_{\rm
eff}$ and photometric colors are too small to alter significantly
the correspondence between theoretical and empirical colors as
well as temperatures discussed in Sect.~\ref{teffscales}.
Nevertheless, our calculations provide an estimate of the
intrinsic limitation of 1D hydrostatic model atmospheres in
reproducing the radiative properties of evolved late-type giants.

In convective stellar atmosphere models one often encounters the
situation that the temperature gradient in the continuum forming
layers depends to some extent on the mixing length parameter. This
has the consequence that predicted stellar colors depend on this
parameter which also allows to offset mismatches to observed
colors by choosing a suitable value \citep[e.g.,][]{HKV02}. As
stated before, due to the insensitivity to the mixing length
parameter, this possibility does not exist in the case of
late-type giants.  In principle, this makes them interesting test
cases for convection theories. However, due to uncertainties in
our knowledge of other physical properties of their atmospheres --
in particular opacities -- such tests are somewhat hampered in
practice.

\addtocounter{table}{+1}
\begin{table*}[p]
\caption[]{A sample of late-type giants in the solar neighborhood
with precise interferometric derivations of $T_{\rm eff}$ from the
literature. Photometric colors are given in Johnson-Cousins-Glass
system (see Sect.~\ref{sample} for details). {\it References for
spectral types, angular diameters, effective temperatures:} 1.
\citet{vB99}; 2. \citet{P98}; 3. \citet{DBT98}; 4. \citet{DBBR96};
5. \citet{DB93}; 6. \citet{R80}. {\it References for photometry:}
10. SIMBAD; 11. \citet{NL69}; 12. \citet{B02}; 13. \citet{G99};
14. \citet{MM78}; 15. \citet{D02}. \label{samplestars}}
\scriptsize \setlength\tabcolsep{5.0pt}
\begin{tabular}{rrll r@{$\,\pm\,$}l r@{$\,\pm\,$}l ccccccl}
\hline \noalign{\smallskip}
     HD &      HR   & Other              & Sp type   & \multicolumn{2}{c}{$\theta_{\rm R}$} & \multicolumn{2}{c}{$T_{\rm eff}$} & Ref & $V$ & $B\!-\!V$ & $V\!-\!I$ & $V\!-\!K$ & $J\!-\!K$ & Ref \\
        &           &                    &           & \multicolumn{2}{c}{mas}          & \multicolumn{2}{c}{K}             &     &     &           &           &           &           &     \\
\noalign{\smallskip} \hline \noalign{\smallskip}
   1632 &      79   &                    & K5 III    & $2.51$ & $0.05$  & $3844$ & $90$  & 1 &  5.79  &  1.60  &  2.13  &  3.98  &  0.99  &  10, 11, 12   \\
   5316 &     259   &                    & M4 III    & $3.67$ & $0.10$  & $3388$ & $70$  & 1 &  6.23  &  1.62  &  3.21  &  5.19  &  1.18  &  10, 11, 13   \\
   5820 &     284   & WW Psc             & M2.4 III  & $3.16$ & $0.16$  & $3480$ & $90 $ & 6 &  6.11  &  1.67  &  1.97  &  4.53  &  1.05  &  10, 14       \\
   6860 &     337   & $\beta$ And        & M0+IIIa   & $12.5$ & $0.6$   & $4002$ & $178$ & 3 &  2.05  &  1.57  &  1.76  &  3.87  &  1.01  &  10, 14       \\
        &           &                    & M0 III    & $14.35$ & $0.19$ & $3705$ & $45$  & 5 &        &        &        &        &        &               \\
   7087 &     351   & $\chi$ Psc         & G8.5 III  & $1.64$ & $0.07$  & $5087$ & $256$ & 1 &  4.66  &  1.03  &  1.30  &  2.30  &  --    &  10, 14       \\
   8126 &     389   & 91 Psc             & K5 III    & $2.22$ & $0.06$  & $4144$ & $108$ & 1 &  5.23  &  1.39  &  1.80  &  3.29  &  --    &  10, 11       \\
   9640 &     450   &                    & M2 III    & $2.42$ & $0.05$  & $3825$ & $96$  & 1 &  5.91  &  1.50  &  2.18  &  4.04  &  --    &  10, 11       \\
  11928 &     564   &                    & M2 III    & $2.96$ & $0.06$  & $3656$ & $72$  & 1 &  5.83  &  1.58  &  2.49  &  4.38  &  --    &  10, 11, 13   \\
  12929 &     617   & $\alpha$ Ari       & K2--IIIab & $ 6.0$ & $0.6$   & $4790$ & $298$ & 3 &  2.01  &  1.15  &  1.16  &  2.66  &  0.75  &  12, 14       \\
  13325 &     631   & 15 Ari, AV Ari     & M3 III    & $3.75$ & $0.11$  & $3605$ & $84$  & 1 &  5.70  &  1.64  &  2.78  &  4.77  &  1.17  &  10, 11, 13   \\
  15656 &     736   & 14 Tri             & K5 III    & $2.60$ & $0.11$  & $4057$ & $124$ & 1 &  5.16  &  1.47  &  1.87  &  3.52  &  0.84  &  10, 11, 12   \\
  18449 &     882   & 24 Per             & K2 III    & $2.08$ & $0.07$  & $4416$ & $192$ & 1 &  4.93  &  1.24  &  1.52  &  2.83  &  0.75  &  10, 12       \\
  18884 &     911   & $\alpha$ Cet       & M1.5 IIIa & $11.9$ & $0.4$   & $3869$ & $161$ & 3 &  2.53  &  1.64  &  1.97  &  4.16  &  1.06  &  10, 14       \\
  27348 &    1343   & 54 Per             & G8 III    & $1.44$ & $0.13$  & $4878$ & $290$ & 1 &  4.93  &  0.94  &  1.27  &  2.17  &  --    &  10, 11       \\
  28305 &    1409   & $\epsilon$ Tau     & G9.5 III  & $2.57$ & $0.06$  & $5027$ & $280$ & 1 &  3.54  &  1.01  &  1.23  &  2.19  &  0.59  &  10, 14       \\
  29139 &    1457   & $\alpha$ Tau       & K5 III    & $20.44$ & $0.11$ & $3947$ & $41$  & 2 &  0.86  &  1.54  &  1.70  &  3.66  &  0.97  &  10, 14       \\
        &           &                    &           & $20.21$ & $0.30$ & $3970$ & $49$  & 5 &        &        &        &        &        &               \\
  30504 &    1533   & 1 Aur              & K3.5 III  & $2.93$ & $0.08$  & $4067$ & $111$ & 1 &  4.89  &  1.44  &  1.81  &  3.63  &  --    &  10, 11, 13   \\
  30834 &    1551   & 2 Aur              & K2.5 III  & $2.79$ & $0.06$  & $4290$ & $183$ & 1 &  4.77  &  1.41  &  1.87  &  3.28  &  --    &  10           \\
  33463 &           & NSV16257           & M2 III?   & $2.87$ & $0.06$  & $3623$ & $122$ & 1 &  6.40  &  1.80  &  2.54  &  4.79  &  --    &  10, 11       \\
  34559 &    1739   & 109 Tau            & G8 III    & $1.42$ & $0.13$  & $4887$ & $362$ & 1 &  4.92  &  0.94  &  1.13  &  2.03  &  --    &  10, 11       \\
  38656 &    1995   & $\tau$ Aur         & G8 III    & $1.97$ & $0.08$  & $4616$ & $344$ & 1 &  4.53  &  0.94  &  1.22  &  2.19  &  0.58  &  10           \\
  39003 &    2012   & $\nu$ Aur          & K0 III    & $2.79$ & $0.06$  & $4496$ & $329$ & 1 &  3.97  &  1.14  &  1.38  &  2.52  &  0.70  &  10, 12       \\
  42471 &    2189   &                    & M2 III    & $2.94$ & $0.06$  & $3954$ & $205$ & 1 &  5.78  &  1.66  &  2.38  &  4.29  &  --    &  10, 11       \\
  43039 &    2219   & $\kappa$ Aur       & G9 III    & $2.16$ & $0.09$  & $4660$ & $276$ & 1 &  4.34  &  1.02  &  1.33  &  2.40  &  0.65  &  10           \\
  54716 &    2696   & 63 Aur             & K3.5 III  & $2.92$ & $0.10$  & $3995$ & $110$ & 1 &  4.94  &  1.45  &  1.99  &  3.42  &  0.85  &  10, 11, 12   \\
  57669 &    2805   & 66 Aur             & K1 III    & $1.94$ & $0.17$  & $4375$ & $289$ & 1 &  5.22  &  1.24  &  1.35  &  2.72  &  --    &  10, 11       \\
  61338 &    2938   & 74 Gem, NSV3671    & M0.0 III  & $2.97$ & $0.29$  & $4090$ & $200$ & 6 &  5.05  &  1.62  &  1.64  &  3.94  &  0.97  &  11, 12       \\
  78712 &    3639   & RS Cnc             & M6.9 III  & $14.77$ & $0.09$ & $3110$ & $117$ & 2 &  5.95  &  1.67  &  --    &  7.56  &  1.27  &  10, 14       \\
  80493 &    3705   & $\alpha$ Lyn       & K7 IIIab  & $7.4$ & $0.6$    & $3969$ & $220$ & 3 &  3.13  &  1.55  &  1.67  &  3.73  &  1.00  &  10, 14       \\
        &           &                    & K7 III    & $7.98$ & $0.31$  & $3791$ & $82$  & 5 &        &        &        &        &        &               \\
  86663 &    3950   & $\pi$ Leo, NSV4699 & M1.7 IIIab& $4.88$ & $0.28$  & $3710$ & $110$ & 6 &  4.70  &  1.60  &  1.97  &  4.18  &  1.03  &  13, 14       \\
  87837 &    3980   & 31 Leo             & K4.2 III  & $3.55$ & $0.22$  & $3930$ & $120$ & 6 &  4.38  &  1.44  &  1.48  &  3.33  &  0.86  &  12, 13, 14   \\
  99998 &    4432   & 87 Leo             & K4.5 III  & $3.71$ & $0.35$  & $3720$ & $170$ & 6 &  4.77  &  1.56  &  1.55  &  3.57  &  1.00  &  11, 12       \\
 108849 &           & BK Vir             & M7.4 III  & $11.11$ & $0.23$ & $2944$ & $34$  & 2 &  7.28  &  1.54  &  --    &  8.12  &  1.37  &  10, 11, 13   \\
 112300 &    4910   & $\delta$ Vir       & M3+III    & $10.0$ & $0.6$   & $3783$ & $182$ & 3 &  3.38  &  1.59  &  2.22  &  4.59  &  1.07  &  10, 14       \\
 114961 &           & SW Vir             & M8 III    & $17.40$ & $0.12$ & $2921$ & $110$ & 2 &  6.85  &  1.67  &  --    &  8.65  &  1.31  &  10, 11, 13, 15   \\
 119149 &    5150   & 82 Vir             & M2.1 IIIa & $4.34$ & $0.25$  & $3690$ & $110$ & 6 &  5.01  &  1.63  &  1.98  &  4.33  &  1.02  &  11, 13, 15   \\
 120819 &    5215   & NSV6468            & M2 III    & $2.50$ & $0.05$  & $3823$ & $83$  & 1 &  5.87  &  1.62  &  2.20  &  4.18  &  1.04  &  11, 15       \\
 124897 &    5340   & $\alpha$ Boo       & K1.5 III  & $20.91$ & $0.08$ & $4321$ & $44$  & 2 & --0.05 &  1.23  &  1.27  &  2.91  &  0.84  &  10, 14       \\
        &           &                    &           & $19.5$ & $1.0$   & $4628$ & $210$ & 3 &        &        &        &        &        &               \\
        &           &                    &           & $19.5$ & $1.0$   & $4628$ & $133$ & 4 &        &        &        &        &        &               \\
        &           &                    & K1 III    & $20.95$ & $0.2$  & $4294$ & $30$  & 5 &        &        &        &        &        &               \\
 126327 &           & RX Boo             & M8 III    & $18.87$ & $0.12$ & $2786$ & $46$  & 2 &  7.96  &  1.29  &  --    &  9.76  &  0.99  &  10, 11, 13   \\
 127665 &    5429   & $\rho$ Aur, NSV6697 & K3 III    & $3.80$ & $0.12$  & $4440$ & $228$ & 1 &  3.59  &  1.30  &  --    &  2.94  &  0.77  &  10, 12       \\
 130084 &    5510   &                    & M1 III    & $2.07$ & $0.05$  & $3962$ & $108$ & 1 &  6.26  &  1.58  &  2.10  &  4.21  &  1.03  &  10, 11       \\
 131873 &    5563   & $\beta$ UMi, NSV6846 & K4--III   & $9.9$ & $0.8$    & $4086$ & $225$ & 3 &  2.08  &  1.47  &  1.47  &  3.31  &  0.79  &  10, 14       \\
 133774 &    5622   & $\nu$ Lib          & K4.8 III  & $2.85$ & $0.41$  & $3930$ & $270$ & 6 &  5.20  &  1.61  &  1.60  &  3.72  &  1.00  &  11, 12       \\
 134320 &    5638   & 46 Boo             & K2 III    & $1.44$ & $0.06$  & $4532$ & $135$ & 1 &  5.68  &  1.24  &  1.58  &  3.00  &  --    &  10           \\
 135722 &    5681   & $\delta$ Boo, NSV7002 & G8 III    & $2.71$ & $0.06$  & $4994$ & $257$ & 1 &  3.49  &  0.95  &  1.37  &  2.27  &  0.65  &  10, 14       \\
 136512 &    5709   & $o$ CrB, NSV7032   & K0 III    & $1.18$ & $0.07$  & $4757$ & $265$ & 1 &  5.50  &  1.02  &  --    &  2.37  &  --    &  10           \\
 137853 &    5745   & NSV20317           & M1 III    & $2.42$ & $0.05$  & $3833$ & $85$  & 1 &  6.04  &  1.60  &  2.20  &  4.20  &  1.01  &  10, 11, 13   \\
 139216 &           & $\tau^{4}$ Ser     & M7.4 III  & $11.60$ & $0.18$ & $3034$ & $30$  & 2 &  6.53  &  1.43  &  --    &  7.53  &  1.23  &  10, 11, 13   \\
 139663 &    5824   & 42 Lib             & K3 III:   & $2.44$ & $0.33$  & $4000$ & $260$ & 6 &  4.96  &  1.34  &  1.29  &  2.88  &  0.82  &  12, 14       \\
 146051 &    6056   & $\delta$ Oph, NSV7556 & M0.5 III  & $10.43$ & $0.48$ & $3779$ & $96$  & 2 &  2.75  &  1.59  &  1.82  &  3.94  &  1.00  &  10, 14       \\
        &           &                    &           & $9.5$ & $0.4$    & $3987$ & $168$ & 3 &        &        &        &        &        &               \\
        &           &                    &           & $9.5$ & $0.5$    & $3983$ & $117$ & 4 &        &        &        &        &        &               \\
 147749 &    6107   & $\nu^{1}$ CrB, NSV7676 & M2 III    & $3.68$ & $0.10$  & $3764$ & $136$ & 1 &  5.20  &  1.60  &  2.45  &  4.33  &  1.06  &  10, 11, 13   \\
 150580 &    6208   &                    & K3        & $1.24$ & $0.06$  & $4555$ & $139$ & 1 &  6.07  &  1.31  &  1.50  &  3.12  &  --    &  10, 11       \\
 152173 &    6258   & 50 Her, NSV20792   & M1 III    & $2.32$ & $0.06$  & $4134$ & $124$ & 1 &  5.72  &  1.61  &  2.09  &  3.89  &  1.00  &  10, 11, 13, 15  \\
 160677 &    6584   &                    & M2 III    & $2.26$ & $0.05$  & $3911$ & $76$  & 1 &  6.06  &  1.58  &  2.28  &  4.22  &  1.06  &  10, 11, 13, 15 \\
 164058 &    6705   & $\gamma$ Dra       & K5 III    & $9.8$ & $0.3$    & $4095$ & $163$ & 3 &  2.22  &  1.52  &  1.55  &  3.55  &  0.95  &  10, 14       \\
        &           &                    &           & $9.8$ & $0.3$    & $4099$ & $80$  & 4 &        &        &        &        &        &               \\
        &           &                    &           & $10.13$ & $0.24$ & $3981$ & $62$  & 5 &        &        &        &        &        &               \\
 167193 &    6820   &                    & K4 III    & $1.54$ & $0.06$  & $4080$ & $99$  & 1 &  6.12  &  1.47  &  1.73  &  3.48  &  --    &  10, 11       \\
 169916 &    6913   & $\lambda$ Sgr      & K2 III:   & $4.29$ & $0.33$  & $4660$ & $180$ & 6 &  2.82  &  1.05  &  1.04  &  2.41  &  0.66  &  12, 14       \\
 175775 &    7150   & $\xi^2$ Sgr        & K1 III:   & $3.79$ & $0.41$  & $4230$ & $230$ & 6 &  3.52  &  1.18  &  1.10  &  2.57  &  0.71  &  12, 14       \\
 177808 &    7237   &                    & M0 III    & $2.27$ & $0.05$  & $4075$ & $86$  & 1 &  5.56  &  1.54  &  1.92  &  3.73  &  0.99  &  10, 11, 12, 13 \\
 177809 &    7238   & NSV24682           & M2.5 III  & $2.40$ & $0.05$  & $3859$ & $78$  & 1 &  6.06  &  1.55  &  2.33  &  4.33  &  1.08  &  10, 11, 13, 15 \\
 189319 &    7635   & $\gamma$ Sge, NSV12638 & M0--III   & $5.6$ & $0.5$    & $4189$ & $238$ & 3 &  3.47  &  1.57  &  1.66  &  3.62  &  0.95  &  10, 14       \\
 194317 &    7806   & 39 Cyg             & K2.5 III  & $2.99$ & $0.08$  & $4192$ & $239$ & 1 &  4.44  &  1.33  &  1.72  &  2.99  &  0.76  &  10, 13       \\
 196610 &    7886   & EU Del             & M6 III    & $11.02$ & $0.25$ & $3243$ & $79$  & 2 &  6.05  &  1.46  &  --    &  7.09  &  1.19  &  10, 11, 13   \\
\noalign{\smallskip} \hline
\end{tabular}
\end{table*}

\addtocounter{table}{-1}
\begin{table*}
\caption{Continued.} \scriptsize \setlength\tabcolsep{5.0pt}
\begin{tabular}{rrll r@{$\,\pm\,$}l r@{$\,\pm\,$}l ccccccl}
\hline \noalign{\smallskip}
     HD &      HR   & Other              & Sp type   & \multicolumn{2}{c}{$\theta_{\rm R}$} & \multicolumn{2}{c}{$T_{\rm eff}$} & Ref & $V$ & $B\!-\!V$ & $V\!-\!I$ & $V\!-\!K$ & $J\!-\!K$ & Ref \\
        &           &                    &           & \multicolumn{2}{c}{mas}          & \multicolumn{2}{c}{K}             &     &     &           &           &           &           &     \\
\noalign{\smallskip} \hline \noalign{\smallskip}
 196777 &    7900   & $\upsilon$ Cap, NSV25208 & M2.1 III  & $4.72$ & $0.52$  & $3540$ & $190$ & 6 &  5.10  &  1.66  &  1.96  &  4.29  &  1.07  &  11, 13, 15   \\
 199169 &    8008   & 32 Vul, NSV13398   & K4 III    & $2.46$ & $0.05$  & $4324$ & $113$ & 1 &  5.01  &  1.48  &  1.82  &  3.40  &  0.87  &  10, 11, 13, 15 \\
 200044 &    8044   & NSV13454           & M3 III    & $3.33$ & $0.08$  & $3615$ & $77$  & 1 &  5.65  &  1.61  &  2.58  &  4.42  &  --    &  10, 11       \\
 212988 &    8555   &                    & K3        & $1.52$ & $0.06$  & $4507$ & $123$ & 1 &  5.98  &  1.45  &  1.64  &  3.35  &  --    &  10, 11       \\
 216386 &    8698   & $\lambda$ Aqr      & M2.5 III  & $9.1$ & $0.7$    & $3477$ & $187$ & 3 &  3.79  &  1.65  &  2.04  &  4.47  &  1.12  &  10, 14       \\
        &           &                    &           & $9.1$ & $1.0$    & $3477$ & $200$ & 4 &        &        &        &        &        &               \\
        &           &                    & M2.0 IIIa & $8.21$ & $0.44$  & $3750$ & $100$ & 6 &        &        &        &        &        &               \\
 219215 &    8834   & $\phi$ Aqr, NSV26044 & M1.5 III  & $5.44$ & $0.89$  & $3770$ & $300$ & 6 &  4.22  &  1.55  &  1.85  &  4.03  &  1.00  &  14           \\
 221345 &    8930   & 14 And             & K0 III    & $1.79$ & $0.07$  & $4206$ & $165$ & 1 &  5.22  &  1.03  &  1.32  &  2.49  &  --    &  10, 11       \\
 221662 &    8942   & NSV26103           & M3 III    & $3.70$ & $0.10$  & $3542$ & $103$ & 1 &  6.06  &  1.71  &  2.97  &  5.05  &  --    &  10, 11       \\
 223755 &    9035   &                    & M2.5 III  & $2.46$ & $0.05$  & $3660$ & $88$  & 1 &  6.12  &  1.60  &  1.93  &  4.09  &  --    &  10, 11       \\
 224303 &    9055   & NSV26170           & M2 III    & $2.41$ & $0.05$  & $3704$ & $80$  & 1 &  6.16  &  1.60  &  2.22  &  4.18  &  --    &  10, 11       \\
 224427 &    9064   & $\psi$ Peg, NSV14777 & M3 III    & $6.5$ & $0.6$    & $3475$ & $206$ & 3 &  4.66  &  1.59  &  2.18  &  4.61  &  1.08  &  10, 14       \\
\noalign{\smallskip} \hline
\end{tabular}
\end{table*}

\section{Synthetic colors versus observations: sample of late-type giants for the comparisons\label{sample}}

\subsection{Selection criteria and stellar sample\label{sample1}}

In order to make a meaningful comparison of the calculated
synthetic colors with observations, a sample of stars is required
for which precise basic stellar parameters (effective temperature,
metallicity, gravity) are known. Ideally, such a sample should
span the entire parameter range in $T_{\rm eff}$, $\log g$, $\MoH$
covered by the grid of synthetic colors and should be complemented
with reliable estimates of interstellar extinction for individual
stars. Unfortunately, this is very difficult to achieve in
practice.

A direct estimate of $T_{\rm eff}$ is possible if the angular
diameter of a star is known (from interferometric measurements or
lunar occultations, for instance). Recent advances in stellar
interferometry and the increasing number of interferometric setups
operating at optical (Mark III, NPOI) and near-infrared (IOTA,
PTI) wavelengths, together with observations of lunar
occultations, now allow the determination of precise angular
diameters and effective temperatures for a number of late-type
giants in the solar neighborhood \citep[e.g.,][]{RP02}. Since
these stars are bright, they are generally well observed, with a
wealth of supplementary photometric and spectroscopic data readily
available, and could therefore form the basis for a sample to be
used for the purposes of this study.

A sample of late-type giants was thus selected according to the
following criteria: (i) the star is a normal giant of spectral
class G, K or M (no chemically or otherwise peculiar stars are
included); (ii) interferometrically derived $T_{\rm eff}$ is
available from the literature ($T_{\rm eff}\leq 5000$\,K); (iii)
the precision of the derived $T_{\rm eff}$ is better than $8\%$
(note that the accuracy is in general significantly better -- see
below); (iv) the star is not variable, or the amplitude of
variability is less than $\Delta V\sim0.1$. A search through the
literature resulted in 56 objects matching these criteria.

Additionally, we included non-variable late-type giants from a
sample of Ridgway et al. (1980; R80), for which effective
temperatures were derived from lunar occultations. Historically,
the R80 effective temperature scale was an important step towards
a homogeneous $T_{\rm eff}$--$(V-K)$ relation for cool stars based
on precise measurements of angular diameters, and it has been
extensively used ever since.

We also included all stars from Perrin et al. (1998; P98), to
illustrate the behavior of $T_{\rm eff}$--color relations at low
effective temperatures. Though these are all variable stars, the
$T_{\rm eff}$--color scale of P98 is the only available which
extends to effective temperatures as low as $T_{\rm
eff}\sim2800$\,K. Data from this sample were not used for the
derivation of $T_{\rm eff}$--$\log g$--color scales, though.

The final sample consists of 74 nearby late-type giants with
precisely derived $T_{\rm eff}$, either from interferometry (Di
Benedetto 1993 (DB93); Dyck et al. 1996 (D96), 1998 (D98); P98,
van Belle et al. 1999 (VB99)), or lunar occultations (R80). All
stars are listed in Table~\ref{samplestars}, along with their
angular diameters (Rosseland diameters, see Sect.~\ref{sample5})
and effective temperatures. The median value of the (reported)
error in $T_{\rm eff}$ for the sample stars is $\pm140$\,K
($\sim$3.5\%); only 7 measurements have errors larger than $6\%$,
while for 53 stars they are less than $5\%$.

\subsection{Broad-band photometric colors\label{obscolors}}

Broad-band \emph{BVIJK} photometric colors of stars in Table~3
were collected from published data using the SIMBAD database. For
the majority of stars data from multiple bibliographical sources
were available, thus averaged colors were used in such cases (68
objects). Apart from a slight inconsistency in $I$-band colors
(see below), colors extracted from different sources agree well,
typically to within $\sim$0.04\,mag or better.

$I$-band photometry was extracted from the SIMBAD database
(Johnson $I$), and/or the Two-Micron Sky Survey (TMSS) catalog of
\citet{NL69}; 20 objects had $I$-band observations in both
systems, while none had photometry in Cousins $I$. The $V-I$
colors were transformed to the Johnson-Cousins-Glass system using
relations from \citet{F83} for Johnson $(V-I)_{\rm J}$ and from
\citet{BW87} for $(V-I)_{\rm TMSS}$ from the TMSS catalog.

\begin{figure}[t]
\includegraphics[width=8.8cm] {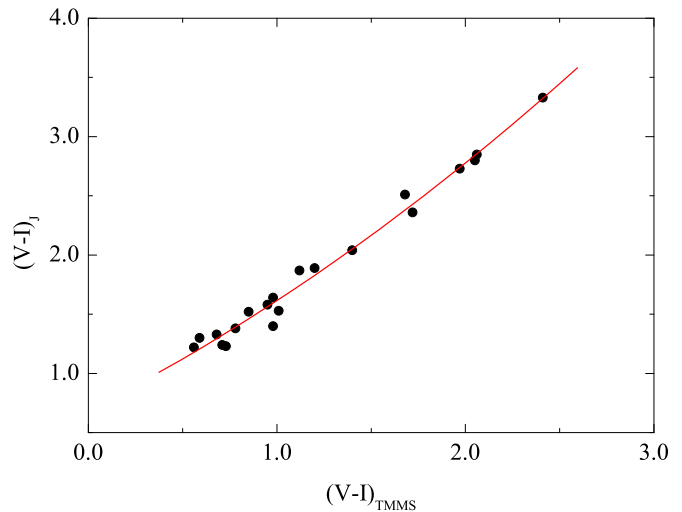}
\caption[]{Johnson $(V-I)_{\rm J}$ vs. $(V-I)_{\rm TMMS}$ from the
Two Micron Sky Survey catalog of \citet{NL69}. Solid line is the
derived relation between $(V-I)_{\rm J}$ and $(V-I)_{\rm TMMS}$
colors (given in numerical form in Eq.~1)\label{photcal}.}
\end{figure}

We find slight systematic differences between the $(V-I)_{\rm C}$
colors resulting from $(V-I)_{\rm J}$ and $(V-I)_{\rm TMSS}$. Note
that \citet{N65} found no systematic discrepancies between their
$I$ magnitudes and those obtained by \citet{K53}, at least to
within an uncertainty margin of about $\pm0.2$ mag. The
discrepancies that we find are most pronounced at bluer colors
(i.e., higher effective temperatures): $\Delta (V-I)\sim 0.20$ at
$(V-I)_{\rm TMSS}\sim0.6$, with Cousins $(V-I)$ colors resulting
from $(V-I)_{\rm TMSS}$ being bluer. The differences gradually
become smaller for redder colors and diminish at $(V-I)_{\rm TMSS}
\ga 2.5$. It should be noted though, that these systematic
differences are always within the margin of $\pm0.2$ mag, which
may possibly explain why they were not reported by \citet{N65}.

Since these trends may suggest the existence of slight differences
in the realization of the TMSS $I$ and standard Kron $I$ passbands
\citep[the latter defined by][]{K57}, we used the following
procedure to obtain Cousins $(V-I)_{\rm C}$. First, stars from our
sample with both $I_{\rm J}$ and $I_{\rm TMSS}$ photometry (20
objects) were used to derive the following relation between
$(V-I)_{\rm J}$ and $(V-I)_{\rm TMSS}$:
\begin{equation}
(V-I)_{\rm J}= 0.690 + 0.806 (V-I)_{\rm TMSS} + 0.1188 (V-I)_{\rm
TMSS}^{2}
\end{equation}
which is valid for $0.5\le (V-I)_{\rm TMSS} \le 2.5$. The RMS
residual of the fit is 0.086 mag (Fig.~\ref{photcal}). This
equation was employed to obtain Johnson $(V-I)_{\rm J}$ colors for
stars with no Johnson $I$-band photometry available. Then, Johnson
$(V-I)_{\rm J}$ colors obtained from $(V-I)_{\rm TMSS}$ and those
originally available in the Johnson system were finally
transformed to Johnson-Cousins-Glass system using the relations
from \citet{F83}.

Near-infrared \emph{JK} colors were converted to the standard
Johnson-Cousins-Glass system using transformations given in
\citet{BB88}.

\emph{BVIJK} colors transformed to the Johnson-Cousins-Glass
system, together with references to the original photometry
sources, are given in Table~\ref{samplestars}.

\subsection{Interstellar reddening}

For most stars in the sample interstellar reddening was derived in
the original interferometry/occultation papers, either from the
difference between intrinsic (for a given spectral type) and
observed broad-band color (D96, D98, VB99), or employing an
empirical model of Galactic extinction (DB93). In both cases, the
derived interstellar reddenings have been found to be small for
the majority of stars; D96, for instance, find a mean color excess
of $E(B-V)\sim0.01$ for all stars in their sample. We thus apply
no corrections for interstellar extinction for the observed
broad-band colors of stars in our sample.

\subsection{Metallicities}

A search through the spectroscopic catalogs available in the
SIMBAD database has yielded metallicities for 37 (out of 74) stars
in our sample. Averaged quantities were used when multiple
derivations of $\FeoH$ were available (19 objects). For the
majority of stars metallicities are close to solar, with a mean
$\FeoH\simeq-0.16$ and standard deviation of $\simeq$0.20\,dex.
Indeed, as we use this sample for the comparison with synthetic
colors at $\FeoH=0.0$, this may result in slight systematic
discrepancies, especially in $T_{\rm eff}-(B-V)$ plane. The
difference between synthetic $B-V$ colors at $\FeoH=0.0$ and
$\FeoH\sim-0.16$ is $\sim$0.03\,mag for $T_{\rm
eff}=3800$--5000\,K ($T_{\rm eff}$--$\log g$ scale of Houdashelt
et al. 2000), with colors at lower metallicity being bluer. The
differences for other colors are considerably smaller (typically
$\sim$0.01\,mag or less). This corresponds to a difference of
$\sim$50\,K in $T_{\rm eff}$ resulting from $B-V$ at $T_{\rm
eff}\geq3800$\,K (less than $\sim$20\,K in other colors).

\begin{figure}[tb]
\includegraphics[width=8.8cm] {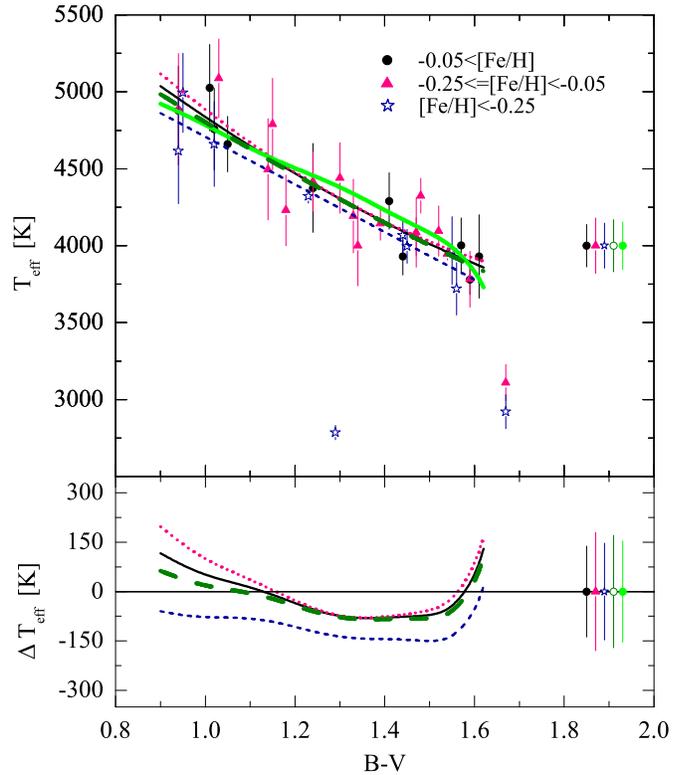}
\caption[]{{\bf Top:} $T_{\rm eff}$--$(B-V)$ diagram for stars
with known metallicities. Different symbols stand for the
following metallicity bins: circles $\FeoH\geq-0.05$ (bin 1),
triangles $-0.25\leq\FeoH<-0.05$ (bin 2), asterisks $\FeoH<-0.25$
(bin 3). Solid, dotted, and short-dashed lines are the best-fits
in the metallicity bins 1--3, respectively. Thick long-dashed line
is a best-fit to all 37 stars with known metallicities, thick
solid line is a best-fit to the entire sample of 74 stars used in
our study (see Sect.~\ref{sample1} for details). A cluster of
symbols with error bars on the right shows the RMS residuals of
the corresponding best-fits. {\bf Bottom:} the difference between
the best-fits in various metallicity bins and the best-fit to the
sample of 74 stars, $\Delta T_{\rm eff}=T_{\rm eff}^{\rm
bin}-T_{\rm eff}^{\rm all}$ (solid, dotted, and short-dashed
lines, for the metallicity bins 1--3 respectively). Thick
long-dashed line shows the difference between the best-fits to the
samples of 37 and 74 stars. RMS residuals of the individual
best-fits are shown on the right\label{feh}.}
\end{figure}

It should be noted though, that we find no clear evidence for the
metallicity effects in the $T_{\rm eff}$--color and color--color
planes, most likely because the spread in metallicity is small
(especially if compared with typical errors in metallicity
determinations, e.g., 0.2--0.3\,dex). This is illustrated in
Fig.~\ref{feh} which shows the $T_{\rm eff}$--$(B-V)$ relation for
37 stars with known $\FeoH$. We divide these 37 stars into three
metallicity bins as indicated in the figure and produce the
quadratic best-fits for each bin, as well as for the entire sample
of 37 stars with available metallicity determinations. The
resulting best-fits are shown as solid lines in the upper panel of
Fig.~\ref{feh} (thick dashed line for the sample of 37 stars). We
also show a best-fit in the $T_{\rm eff}$--$(B-V)$ plane for the
sample containing all 74 stars used in our further analysis (thick
solid line; see Sect.~\ref{teffscales1} for details). The RMS
residuals of the best-fits are 140\,K, 180\,K, and 150\,K in
metallicity bins 1--3 respectively, 150\,K for the sample of 37
stars and 170\,K for all 74 stars. The mean difference between the
best-fit representing the 37 stars and those corresponding to
different metallicity bins are 20\,K, 40\,K and $-$80\,K, for bins
1--3 respectively. Thus only the lowest metallicity bin is
slightly more deviant, however, in all cases the differences are
not statistically significant. Note, however, that the best-fit
representing all 37 stars with metallicity determinations is
slightly deviant from the best-fit which is based on the entire
sample of 74 late-type giants. While the mean difference is small
($-$20\,K, with the best-fit for 37 stars predicting lower $T_{\rm
eff}$ for a given $B-V$), differences in certain effective
temperature ranges may be considerable (e.g., $-$80\,K at $T_{\rm
eff}\sim4000$--4400\,K), though this is again significantly less
than the typical spread of observed $T_{\rm eff}$ in the
individual samples.

The differences between the best-fits in individual metallicity
bins are considerably smaller in other $T_{\rm eff}$--color and
color--color diagrams, with no evident systematical trends.

\subsection{Comparison of angular diameters, fluxes and effective temperatures\label{sample5}}

\begin{figure*}[t]
\sidecaption
\includegraphics[width=13.28cm] {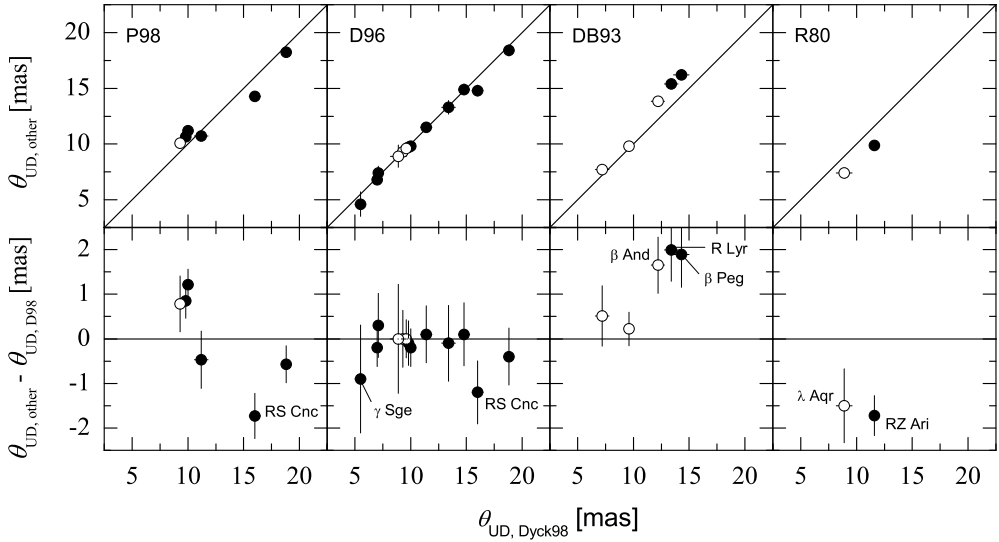}
\caption[]{~Comparison of angular diameters measured by different
authors (see text for details). Open circles are the late-type
giants used in this work.\label{diameters}}
\end{figure*}

Once the angular diameter of a star is known, the effective
temperature can be estimated through the following relation (D96):
\begin{equation}
T_{\rm eff}=1.316\times10^6\, \left( \frac{\Fbol}{\theta_{\rm
R}^2}\right)^{1/4}
\end{equation}
here \thetaR\ is the angular diameter (in mas), and \Fbol\ is the
bolometric flux (in ${\rm W\,m^{-2}}$). Obviously, potential
systematic errors in derived $T_{\rm eff}$ may result both from
the differences in derived stellar radii, $\theta_{\rm R}$, and
observed bolometric fluxes, $\Fbol$.

Bolometric fluxes are usually obtained by integrating observed
photometric fluxes over as large wavelength interval as possible.
After carefully checking the $\Fbol$ given in the original papers
(R80, DB93, D96, D98, P98, VB99), we find no significant
systematic differences in bolometric fluxes obtained by different
groups (exceptions and possible implications are discussed in
Sect.~\ref{sample52}). In the forthcoming sections we will thus
concentrate on the angular diameters and effective temperatures.

\subsubsection{Angular diameters}

The {\em measured} angular diameter of a star (i.e., a uniform
disk diameter, \thetaUD) does not exactly represent the {\em true}
angular diameter and should be corrected for the limb-darkening.
This is done by applying a correction (which varies from 1.00 for
a uniformly bright disk to $\sim$1.13 for a fully darkened disk),
defined as a ratio of the limb-darkened diameter to the uniform
disk diameter (at a given wavelength). The correction is derived
from stellar atmosphere models \citep[see][ for a detailed
discussion]{ST87,S97}.

The limb-darkened diameter in Eq.~(2) is the Rosseland diameter,
which corresponds to a surface where the Rosseland mean optical
depth is equal to unity. As advocated by \citet{ST87}, the
temperature calculated at this surface provides a good estimate of
$T_{\rm eff}$ (see also Baschek et al. 1991 for a more extensive
discussion on the definitions of radii and $T_{\rm eff}$ of red
giants). Effective temperatures of giants in our sample were
derived in the original interferometry papers, employing Rosseland
angular diameters which were calculated using the following
corrections: $\thetaR \simeq 1.022 \thetaUD$ by D96, D98 and VB99;
$\thetaR \simeq 1.035 \thetaUD$ by P98; similar conversion factors
were used for individual stars by DB93 and R80. The Rosseland
angular diameters of the sample giants are given in
Table~\ref{samplestars}.

There have been indications discussed in the literature that
systematical differences may exist between the uniform disk
diameters obtained with different interferometric setups. In their
analysis of the Infrared Optical Telescope Array (IOTA) data, D98
have found indications of systematical differences between
\thetaUD\ obtained with the CERGA interferometer
\citep{DBR87,DBF90} and those obtained with the classical (D98)
and FLUOR (P98) beam combiners at IOTA (all operating in the
near-infrared $K$-band). To the contrary, a recent comparison of
angular diameters obtained with optical interferometers (Mark III
and NPOI) has revealed no systematic differences \citep{N01}.

The largest fraction of stars used in this work comes from the
sample of VB99. However, only three stars in this data set have
previous measurements of angular diameters from occultations and
interferometry, while none of these three stars is available in
the other data sets used in this work. As remarked by VB99,
however, angular diameters and effective temperatures of four
stars in their sample are in good agreement with those inferred
from the infrared flux method. There is also a good agreement with
an upper limit estimate for \thetaUD\ obtained for one of the
stars (HR 2630) from lunar occultation. For the two stars with
interferometric measurements of \thetaUD\ the discrepancies
are larger, apparently due to large errors in earlier derivations
of \thetaUD\ (see VB99 for a detailed discussion).

A more straightforward comparison can be made for stars from other
samples. In Fig.~\ref{diameters} we compare angular diameters from
D98 with those derived by DB93, P98, D96, and R80. Stars used in
this work are indicated by open circles.

A good agreement can be seen between the samples of D98 and D96;
in both cases stars were measured using the same classical beam
combiner at IOTA interferometer. There is also a good consistency
between angular diameters measured by D98 and P98, who used a
FLUOR fiber beam combiner at IOTA; the data scatter however is
large. One star, RS Cnc, deviates significantly; however, it is a
semiregular variable (of type SRC) with a photographic amplitude
of $\sim$1.5\,mag \citep[][ GCVS]{GCVS}, thus the discrepancy may
also reflect a real variation of stellar diameter. Interestingly,
the angular diameter of this star obtained by D96 is also
different from that derived by D98 (note that this star is not in
the sample used in our study).

The two stars in the sample of R80, RZ Ari and $\lambda$ Aqr, have
significantly lower angular diameters than measured by D98. The
former is a semiregular variable star of type SRB, with an
amplitude of $\Delta V\sim0.4$ (GCVS), while the latter is an
irregular variable of type LB, with $\Delta V\sim0.1$ (GCVS); at
least in the latter case variability alone can not account for the
discrepancy.

As it was already indicated by D98, angular diameters measured by
DB93 are indeed systematically larger than those obtained by D98
(Fig.~\ref{diameters}). There is also a tendency that these
differences increase with increasing angular diameter of a star
(Fig.~\ref{diameters}).

There are two stars in the sample of DB93 which have measurements
of angular diameters available from other groups. One star ($\mu$
Gem) was observed by R80; the DB93 value \citep[originaly measured
by][]{DBR87}, $13.50\pm0.15$\,mas, is indeed larger than that
obtained by R80, $12.10\pm0.39$\,mas. The other star is $\alpha$
Tau, which was measured by P98. In this case the obtained angular
diameters agree well, with $19.60\pm0.29$\,mas obtained by DB93
and $19.75\pm0.11$\,mas by P98.

Obviously, the largest systematic discrepancies arise between the
data sets of DB93 and D98; angular diameters of five stars
obtained by DB93 tend to be larger than those derived by D98. The
measured radii of the two stars common to D98 and R80 are rather
discrepant too, as are the measurements of RS Cnc by D98 and P98.
Note that the scatter of the individual measurements in
Fig.~\ref{diameters} is generally large. This is perhaps even more
surprising, as the majority of these stars are non-variable and in
some cases their angular diameters are measured using the same
instrument.

\begin{figure*}[t]
\sidecaption
\includegraphics[width=13cm]{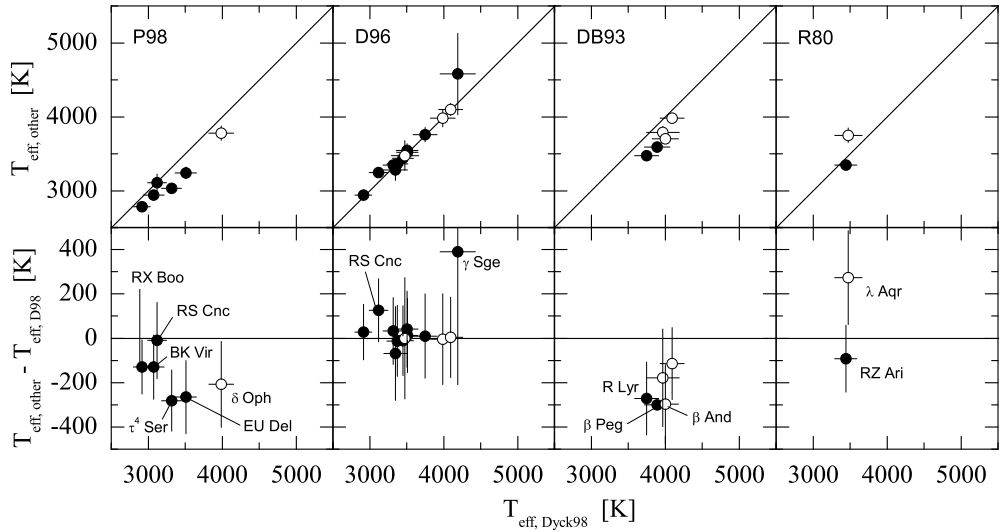}
\caption[]{~Comparison of effective temperatures measured by
different authors (see text for details). Open circles are the
late-type giants used in this work.\label{teffs}}
\end{figure*}

While pinning down the precise source of these discrepancies is
beyond the scope of this study, several comments may be
appropriate. First, it should be noted that angular diameters
quoted in DB93 are taken from the earlier study of \citet{DBR87},
with the original interferometric measurements obtained more than
two decades ago. The angular diameters derived in \citet{DBR87}
are thus amongst the pioneering interferometric measurements of
the angular diameters of late-type giants, with typically large
uncertainties in the data reduction procedure (calibration,
conversion from the uniform to limb-darkened angular diameters,
and so forth). Second, angular diameters in D98 were mostly
derived from a single observation of the visibility made at one
spatial frequency point, which clearly may introduce additional
systematic uncertainties. For instance, according to \citet{WAK04}
visibility function of the late-type giant $\psi$ Phe measured
with the VLTI/VINCI is significantly different from both the
uniform disk and fully-darkened disk models, but it seems that
this discrepancy is clearly discerned only in the second lobe of
the visibility function.

Indeed, stars that are common to different data sets used in our
study are too few to decide firmly whether the discrepancies
mentioned above point towards the general systematic differences,
or they simply show the scatter due to small number statistics.
Nevertheless, all these differences clearly indicate that the real
errors in the measured stellar radii may be considerably larger
than indicated by the error bars provided in the individual
studies. Inevitably, this has a direct effect on the effective
temperatures derived using the observed stellar radii, thus
putting a lower limit on the data scatter in observed $T_{\rm
eff}$--color relations.

\subsubsection{Effective temperatures\label{sample52}}

Effective temperatures calculated in the original interferometry
papers (R80, DB93, D96, D98, P98, VB99) are given in
Table~\ref{samplestars}. A comparison of $T_{\rm eff}$ derived by
different groups is shown in Fig.~\ref{teffs}, with stars used in
this work indicated by open circles.

As in the case of angular diameters, the agreement in $T_{\rm
eff}$ derived by D98 and D96 is very good. It should be mentioned
though, that identical values of $\Fbol$ were used in both
studies, thus the consistency in derived $T_{\rm eff}$ simply
reflects the agreement in derived angular diameters.

Effective temperatures derived by P98, however, tend to be
systematically lower than those obtained by D98. This is
determined either by a larger angular diameter ($\delta$ Oph),
smaller $\Fbol$ (RX Boo, BK Vir), or both (EU Del, $\tau^{4}$
Ser), as derived by P98. While both angular diameter and $\Fbol$
are significantly deviating in case of RS Cnc, both quantities are
smaller than those derived by D98 by a similar factor, thus
leaving the resulting $T_{\rm eff}$ unaltered.

Effective temperatures derived by DB93 also tend to be
systematically lower than those of D98. This is a consequence of
systematically larger angular diameters of DB93. The $T_{\rm eff}$
of $\lambda$ Aqr derived by R80 is significantly higher ($\Delta
T_{\rm eff}\sim 270$\,K) than that obtained by D98, which is a
consequence of considerably smaller angular diameter derived by
R80.

The discrepancies in $T_{\rm eff}$ derived by various groups are
thus non-negligible, and result from the differences both in
angular diameters (predominantly) and bolometric fluxes. Effective
temperatures of stars in the samples of P98 and DB93 seem to be
systematically lower than those obtained by D98. A comparison of
general trends of stars from different samples in the $T_{\rm
eff}$--color diagrams gives an indication that the $T_{\rm eff}$
derived by D98 are somewhat higher ($\sim$50\,K) than the average
trend of the sample including all 74 stars, while the effective
temperatures of P98 and DB93 tend to be lower by a comparable
amount.

To investigate a possible effect of these systematic differences
on the $T_{\rm eff}$--color and color--color relations derived in
Sect.~\ref{teffscales}, we produced $T_{\rm eff}$--color scales in
different color planes without including stars from the sample of
D98 (while using all stars from the other samples). This procedure
was repeated by excluding each sample in turn from the whole
sample of 74 stars. In all cases, the resulting $T_{\rm
eff}$--color relations were essentially unaltered. Typically, the
differences between the $T_{\rm eff}$--color relations based on
all 74 stars and those with stars from a certain sample excluded
were within $\sim$70\,K for $T_{\rm eff}=3700$--4800\,K,
suggesting that the influence of these systematical differences on
the derived $T_{\rm eff}$--color relations is small.

\section{Synthetic photometric colors versus observations: results and discussion\label{teffscales}}

Since precise effective temperatures and broad-band photometric
colors of late-type giants in the solar neighborhood are readily
available, they can be supplemented with the $T_{\rm eff}$--$\log
g$ relation to construct an empirical $T_{\rm eff}$--$\log
g$--color scale based on the observed quantities of late-type
giants (Table~\ref{samplestars}). This scale could be used further
to make a direct comparison of the new synthetic colors with the
observations of late-type giants in our sample, as well as with
available $T_{\rm eff}$--color and color--color relations. It may
also provide a basis for checking the consistency between
synthetic colors calculated with different model atmosphere codes.

\subsection{New $T_{\rm eff}$--$\log g$--color scales\label{teffscales1}}

Although no previous knowledge about the evolutionary status of
stars in Table~\ref{samplestars} is available, it is most likely
that these stars are either on the red giant branch (RGB), or
asymptotic giant branch (AGB). Obviously, their gravities and
effective temperatures have to be related through the appropriate
$T_{\rm eff}$--$\log g$ relations, which should ideally be
different for stars on the RGB and AGB.

\begin{figure}[tb]
\includegraphics[width=8.8cm] {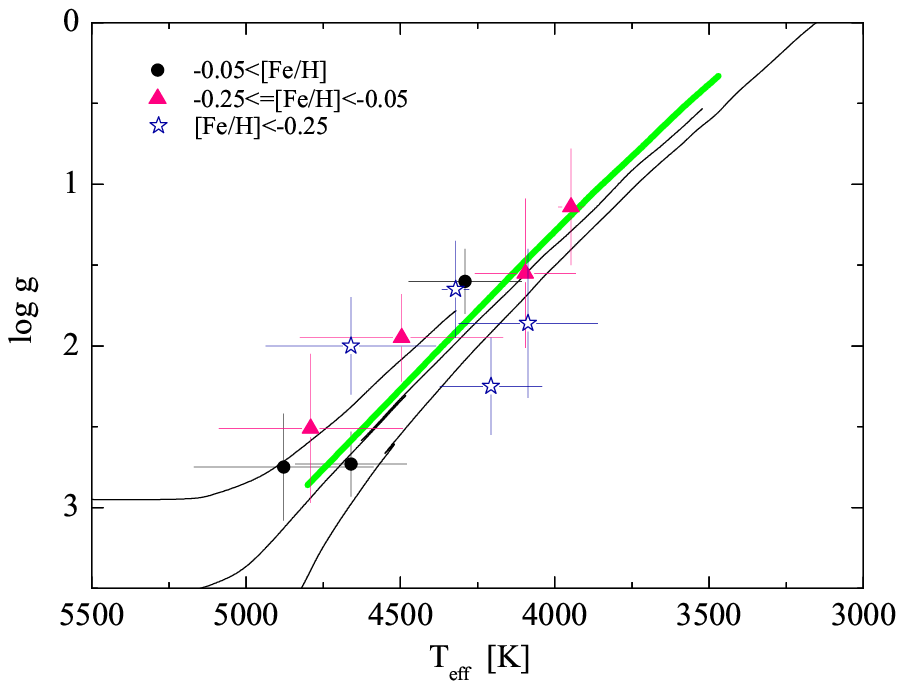}
\caption[]{Stars from Table~\ref{samplestars} with spectroscopic
gravities available from the literature plotted in the $T_{\rm
eff}$--$\log g$ plane. Symbols mark the following metallicity
bins: circles $\FeoH\geq-0.05$, triangles $-0.25\leq\FeoH<-0.05$,
asterisks $\FeoH<-0.25$. Thick solid line corresponds to the
$T_{\rm eff}$--$\log g$ scale of \citet[][ H00]{H00} extrapolated
linearly to $T_{\rm eff}=4800$\,K. Thin lines are isochrones of
\citet{MG01} for 0.5, 2 and 10 Gyr.\label{H00scale}}
\end{figure}

Unfortunately, the present knowledge about the variations of $\log
g$ along the RGB and AGB is rather limited, and relies essentially
on theoretical modeling. Since empirical $\log g$ of RGB and AGB
stars are typically a product of spectroscopic analysis, their
determination becomes extremely complicated at effective
temperatures lower than $\sim$4000\,K (due to problems related
with the definition of continuum, effects of line overlapping,
etc.), uncertainties that grow sharply with decreasing $T_{\rm
eff}$. Effects of metallicity and age may introduce additional
scatter/shifts in the resulting $T_{\rm eff}$--$\log g$ scale.
Though these uncertainties indeed place internal limitations on
the precision of existing $T_{\rm eff}$--$\log g$ calibrations,
more importantly, any empirical $T_{\rm eff}$--$\log g$ scale is
likely to suffer from these uncertainties in a systematical way
too, depending on the properties of the stellar sample from which
it was derived.

Thus, ideally, we would like to construct a $T_{\rm eff}$--$\log
g$ scale which is based entirely on the observed quantities of
stars from our sample. However, only 11 stars from
Table~\ref{samplestars} have spectroscopic gravities available
from the literature. When plotted on the $T_{\rm eff}$--$\log g$
plane (Fig.~\ref{H00scale}), the scatter in the data is too large
to derive a reliable temperature--gravity relation.

For the purposes of this study we thus use one of the existing
empirical $T_{\rm eff}$--$\log g$ relations, namely the scale of
\citet[][ H00]{H00}, which is based on the effective temperatures
derived from interferometry and gravities assigned from
theoretical isochrones (see H00 for details). According to
\citet{H00,H00b}, synthetic broad-band colors based on their
$T_{\rm eff}$--$\log g$ relation agree well with observations of
nearby K and early-M giants in the $T_{\rm eff}$--color planes.
Their relation is also in reasonable agreement with observations
of 11 giants from our sample (Fig.~\ref{H00scale}), although the
scatter in the observed data is indeed large.

\begin{table}[tb]
\caption[]{New empirical $T_{\rm eff}$--$\log g$--color relation
for the late-type giants. The $T_{\rm eff}$--color relations are
best-fits to the observed effective temperatures and colors of 74
late-type giants, while gravities are assigned to a given $T_{\rm
eff}$ according to the $T_{\rm eff}$--$\log g$ relation of
\citet[][ H00]{H00}. The last line gives the RMS residual of the
fit, expressed as a temperature difference. Photometric colors are
in the Johnson-Cousins-Glass system (see Sect.~\ref{teffscales1}
for more details).\label{empscale}}

\centering
\begin{tabular}{cccccccccccccc}
\hline \noalign{\smallskip}
 $T_{\rm eff}$ & $\log g$ & & $B\!-\!V$ & $V\!-\!I$ & $V\!-\!K$ & $J\!-\!K$   \\
\noalign{\smallskip} \hline \noalign{\smallskip}
4800  &  2.90  &                        &  1.021  &  1.016  &  2.280  &  0.639  \\
4700  &  2.69  &                        &  1.075  &  1.054  &  2.392  &  0.675  \\
4600  &  2.49  &                        &  1.133  &  1.096  &  2.514  &  0.712  \\
4500  &  2.28  &                        &  1.195  &  1.144  &  2.649  &  0.750  \\
4400  &  2.08  &                        &  1.263  &  1.198  &  2.798  &  0.789  \\
4300  &  1.89  &                        &  1.338  &  1.261  &  2.963  &  0.828  \\
4200  &  1.69  &                        &  1.408  &  1.334  &  3.146  &  0.869  \\
4100  &  1.50  &                        &  1.469  &  1.421  &  3.350  &  0.910  \\
4000  &  1.30  &                        &  1.528  &  1.526  &  3.580  &  0.952  \\
3900  &  1.11  &                        &  1.573  &  1.648  &  3.839  &  0.996  \\
3800  &  0.93  &                        &  1.605  &  1.786  &  4.133  &  1.040  \\
3700  &  0.74  &                        &  1.622  &  1.962  &  4.469  &  1.086  \\
3600  &  0.57  &                        &  1.628  &  2.192  &  4.855  &  1.133  \\
3500  &  0.39  &                        &  1.630  &  2.501  &  5.304  &  1.181  \\
\noalign{\smallskip} \hline \noalign{\smallskip}
\multicolumn{2}{c}{RMS residual [K]} & &   170   &   160   &   140   &   150   \\
\noalign{\smallskip} \hline \noalign{\smallskip}

\end{tabular}
\end{table}

\begin{table*}[t]
\caption[]{Semi-empirical $T_{\rm eff}$--$\log g$--color relations
for late-type giants based on synthetic colors calculated with
{\tt PHOENIX} and {\tt MARCS} model atmospheres. Photometric
colors are given in the Johnson-Cousins-Glass system (see
Sect.~\ref{teffscales1} for details).\label{sempscale1}}
\centering
\begin{tabular}{ccccccc|cccc}
\hline \noalign{\smallskip}
           &         &     & \multicolumn{4}{c}{\tt PHOENIX} & \multicolumn{4}{c}{\tt MARCS} \\
\cline{4-7} \cline{8-11} \noalign{\smallskip}
 $T_{\rm eff}$ & ${\rm log}\,g$ &  &  $B\!-\!V$ & $V\!-\!I$ & $V\!-\!K$ & $J\!-\!K$ & $B\!-\!V$ & $V\!-\!I$ & $V\!-\!K$ & $J\!-\!K$ \\
\noalign{\smallskip} \hline \noalign{\smallskip}

  4800  &  2.90 & &  1.054  &  1.060  &  2.333  &  0.607  &  1.056  &  1.013  &  2.323  &  0.626  \\
  4700  &  2.69 & &  1.103  &  1.105  &  2.445  &  0.639  &  1.109  &  1.054  &  2.431  &  0.658  \\
  4600  &  2.49 & &  1.150  &  1.152  &  2.562  &  0.673  &  1.164  &  1.100  &  2.547  &  0.693  \\
  4500  &  2.28 & &  1.204  &  1.205  &  2.690  &  0.709  &  1.222  &  1.150  &  2.672  &  0.729  \\
  4400  &  2.08 & &  1.259  &  1.263  &  2.827  &  0.748  &  1.284  &  1.206  &  2.805  &  0.767  \\
  4300  &  1.89 & &  1.315  &  1.329  &  2.975  &  0.789  &  1.345  &  1.267  &  2.950  &  0.808  \\
  4200  &  1.69 & &  1.370  &  1.402  &  3.136  &  0.831  &  1.406  &  1.334  &  3.106  &  0.853  \\
  4100  &  1.50 & &  1.421  &  1.485  &  3.312  &  0.875  &  1.466  &  1.411  &  3.276  &  0.900  \\
  4000  &  1.30 & &  1.473  &  1.587  &  3.512  &  0.921  &  1.529  &  1.506  &  3.468  &  0.947  \\
  3900  &  1.11 & &  1.516  &  1.710  &  3.739  &  0.969  &  1.582  &  1.621  &  3.686  &  0.996  \\
  3800  &  0.93 & &  1.551  &  1.860  &  4.002  &  1.017  &  1.619  &  1.765  &  3.940  &  1.046  \\
  3700  &  0.74 & &  1.573  &  2.056  &  4.321  &  1.064  &  1.636  &  1.950  &  4.245  &  1.096  \\
  3600  &  0.57 & &  1.563  &  2.324  &  4.736  &  1.113  &  1.617  &  2.201  &  4.631  &  1.147  \\
  3500  &  0.39 & &  1.505  &  2.707  &  5.319  &  1.160  &  1.556  &  2.543  &  5.139  &  1.197  \\

\noalign{\smallskip} \hline
\end{tabular}
\end{table*}

We provide several new $T_{\rm eff}$--$\log g$--color scales.
Table~\ref{empscale} gives an empirical $T_{\rm eff}$--$\log
g$--color relation derived using empirical $T_{\rm eff}$--color
relations (thick lines in Fig.~\ref{TCplanes}) obtained as
best-fits to the observed data (stars from
Table~\ref{samplestars}, except those from the sample of P98) and
supplemented with the $T_{\rm eff}$--$\log g$ scale of H00. The
H00 scale used for this purpose was linearly extrapolated to
$T_{\rm eff}=4800$\,K. The typical RMS error of the fitting
procedure is $\sim$160\,K (Table~\ref{empscale}).
Tables~\ref{sempscale1}--\ref{sempscale2} list three additional
semi-empirical scales, based on synthetic colors of {\tt PHOENIX},
{\tt MARCS} and {\tt ATLAS} corresponding to the $T_{\rm
eff}$--$\log g$ relation of H00. All scales are valid for the
effective temperature range of $T_{\rm eff}=3500$--4800\,K. It
should be noted that the H00 scale is representative of RGB stars,
thus appropriate care should be taken when using these new $T_{\rm
eff}$--$\log g$--color relations both at the low and high
effective temperature ends, where RGB stars may be mixed with
objects on the horizontal branch and AGB, respectively.

The scatter in observational data may provide an estimate of the
intrinsic limits in the precision of empirical $T_{\rm
eff}$--color relations, as indicated, for instance, by RMS
residuals of the best fits to the observed sequences of late-type
giants in various $T_{\rm eff}$--color planes. These errors are
typically $\sim$160\,K (Table~\ref{empscale}) and they are
determined by the current uncertainties in interferometrically
derived effective temperatures, various systematical effects,
astrophysical scatter, etc.

\subsection{Comparison of $T_{\rm eff}$--color relations\label{Teff-color}}

\begin{table}[t]
\caption[]{Semi-empirical $T_{\rm eff}$--$\log g$--color relation
for late-type giants based on synthetic colors calculated with
{\tt ATLAS} model atmospheres. Photometric colors are given in the
Johnson-Cousins-Glass system (see Sect.~\ref{teffscales1} for
details).\label{sempscale2}} \centering
\begin{tabular}{ccccccccccc}
\hline \noalign{\smallskip}
        &          &         &          \multicolumn{4}{c}{ATLAS}          \\
\cline{4-7} \noalign{\smallskip}
 $T_{\rm eff}$ & ${\rm log}\,g$ &  & $B\!-\!V$ & $V\!-\!I$ & $V\!-\!K$ & $J\!-\!K$ \\
\noalign{\smallskip} \hline \noalign{\smallskip}

  4750  &  2.79 &  &  1.067  &  1.051  &  2.391  &  0.650  \\
  4500  &  2.28 &  &  1.191  &  1.165  &  2.685  &  0.740  \\
  4250  &  1.79 &  &  1.330  &  1.308  &  3.035  &  0.842  \\
  4000  &  1.30 &  &  1.497  &  1.512  &  3.472  &  0.955  \\
  3750  &  0.84 &  &  1.648  &  1.831  &  4.060  &  1.071  \\
  3500  &  0.39 &  &  1.683  &  2.396  &  4.974  &  1.173  \\

\noalign{\smallskip} \hline
\end{tabular}
\end{table}

Synthetic {\tt PHOENIX}, {\tt MARCS} and {\tt ATLAS} colors given
in Tables~\ref{sempscale1}--\ref{sempscale2} provide three new
semi-empirical $T_{\rm eff}$--color scales which may be readily
compared with the observations of giants from Table~3 in the
$T_{\rm eff}$--color and color--color domains. This also gives a
possibility for a direct comparison between the {\tt PHOENIX},
{\tt MARCS} and {\tt ATLAS} colors, since they are all given for
the same $T_{\rm eff}$--$\log g$ relation of H00. A comparison of
the new relations with observations and published $T_{\rm
eff}$--color scales is given in Fig.~\ref{TCplanes}.

In order to provide a reference frame for discussion of the new
scales, we also include the following recently published $T_{\rm
eff}$--color relations into our analysis:

\begin{itemize}

\item {\em BaSeL 2.2 \citep[][ BaSeL 2.2]{LCB98}:} photometric colors are
derived from theoretical spectra calibrated to match empirical
$T_{\rm eff}$--color relations at $\FeoH=0.0$. BaSeL 2.2 colors
used in this work were calculated for the $T_{\rm eff}$--$\log g$
relation of H00, using the interactive web-based BaSeL server
\footnote{http://tangerine.astro.mat.uc.pt/BaSeL/};

\item {\em \citet[][ A99]{A99}:} empirical scale, obtained using
photometric observations of a large sample of Galactic field and
globular cluster stars at different metallicities. Effective
temperatures of individual stars were derived using the infrared
flux method (IRFM), the best-fits yielding $T_{\rm eff}$--color
and color--color relations;

\item {\em \citet[][ SF00]{SF00}:} empirical $T_{\rm eff}$--$(B-V)$
scale, obtained using observed colors and effective temperatures
of 537 {\it Infrared Space Observatory} (ISO) standard stars from
\citet{DB98}. Effective temperatures of individual stars were
derived from the $T_{\rm eff}$--$(V-K)$ relation, calibrated on a
sample of nearby stars with angular diameters available from
interferometry. For the purposes of this study their $B-V$ colors
were selected according to the $T_{\rm eff}$--$\log g$ relation of
H00;

\item {\em \citet[][ VC03]{VC03}:} empirical scales based on
synthetic \emph{BVRI} colors of \citet{BG78,BG89}, adjusted to
satisfy observational constrains from the CMDs of several Galactic
globular and open clusters, field stars in the solar neighborhood,
empirical $T_{\rm eff}$--color relations, etc. We used VC03 colors
selected according to the $T_{\rm eff}$--$\log g$ scale of H00;

\item {\em \citet[][ H00]{H00}:} theoretical colors, calculated
with MARCS and SSG codes; TiO opacities were adjusted to reproduce
the observed spectra of M giants from \citet{F94}; no ${\rm H_2O}$
opacities were included in the calculations. Their photometric
colors are given for the H00 $T_{\rm eff}$--$\log g$ relation
(provided in the same paper), which was linearly extrapolated by
us to $T_{\rm eff}=4800$\,K;

\item {\em \citet[][ S88]{S88}, \citet[][ Pi98]{Pi98}:} scales
based on photometric colors calculated from the observed spectra
of late-type giants. Note that observed spectra from Pi98 and S88
are compared with those calculated using {\tt PHOENIX} and {\tt
MARCS} model atmospheres in Sect.~\ref{spectra}, thus the two
scales are provided to illustrate the behaviour of photometric
colors in the $T_{\rm eff}$--color planes. Photometric colors of
the S88 and Pi98 scales were calculated by us in the standard
Johnson-Cousins-Glass photometric system using the procedure
described in Sect.~\ref{models3}. Effective temperatures for both
scales were assigned using the effective temperature -- spectral
class relation of Pi98;

\begin{figure*}[tp]
\centering
\includegraphics[width=16cm] {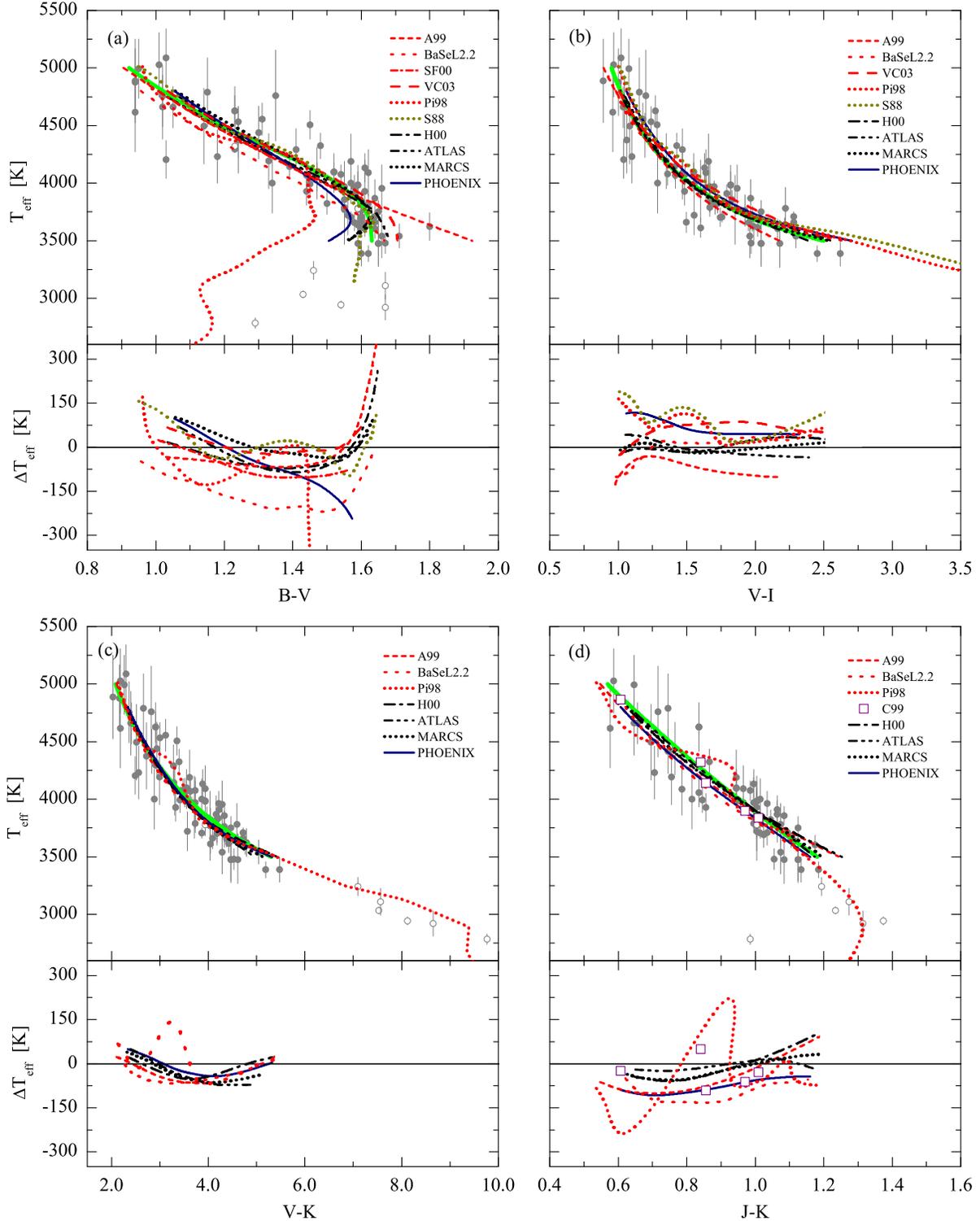}
\caption[]{Empirical and theoretical $T_{\rm eff}$--color
relations for late-type giants in different $T_{\rm eff}$--color
planes (a--d, top panels). Filled circles are late-type giants
from Table~3, stars from the sample of \citet{P98} are highlighted
as open circles. Thick solid line is a best-fit to the data, which
represents a new empirical $T_{\rm eff}$--$\log g$--color scale
(Table~4) in a particular $T_{\rm eff}$--color plane. Several
existing $T_{\rm eff}$--color relations are shown as well,
together with semi-empirical scales constructed using synthetic
colors of {\tt PHOENIX}, {\tt MARCS} and {\tt ATLAS} (Table~5).
Open rectangles in panel (d) are late-type giants from \citet{C99}
(see text for details). Bottom panels in each figure show the
difference between various $T_{\rm eff}$--color relations and the
empirical scale (thick solid line, Table~4) in a given $T_{\rm
eff}$--color plane ($\Delta T_{\rm eff}=T_{\rm eff}^{\rm
other}-T_{\rm eff}^{\rm bestfit}$). \label{TCplanes} }
\end{figure*}

\item {\em \citet[][ C99]{C99}:} photometric colors of the five
late-type giants used by \citet{C99} in the role of templates to
produce a library of near-infrared spectra of late-type giants (a
comparison of several C99 spectra with those produced using
stellar model atmospheres is given in Sect.~\ref{spectra}). The
five late-type giants are: $\beta$\,Gem (K0~III; spectrum taken
from Cohen et al. 1995); $\alpha$\,Boo (K1.5~III; Cohen et al.
1996); $\alpha$\,Hya (K3~III; Cohen et al. 1995); $\alpha$\,Tau
(K5~III; Cohen et al. 1999); and $\beta$\,And (M0~III; Cohen et
al. 1995). Effective temperatures of individual stars were taken
either from Table~\ref{samplestars}, or from \citet{C95, C96,
C99}. Observed $J-K$ colors were collected from the literature
(typically -- SIMBAD database) and converted to the standard
Johnson-Cousins-Glass system using transformations given in
\citet{BB88}.

\end{itemize}

It should be noted, that A99 give an extensive comparison of their
scale with existing $T_{\rm eff}$--color relations; similar
comparisons are also done by SF00 and VC03. Altogether this may
provide a further reference for a comparison of the new scales
derived in this work with other relations published in the
literature.

Generally, the consistency between different $T_{\rm eff}$--color
relations (including those based on synthetic colors of {\tt
PHOENIX}, {\tt MARCS} and {\tt ATLAS}) and observed sequences of
late-type giants (represented by the new empirical scales in
Tables~\ref{sempscale1}--\ref{sempscale2}) is good over a large
range of effective temperatures (or colors). The agreement is
especially good in the $T_{\rm eff}$--$(V-K)$ plane
(Fig.~\ref{TCplanes}c), where {\it all} scales agree to within
$\Delta T_{\rm eff}\sim80$\,K. The only exception is the scale of
Pi98 which predicts considerably higher effective temperatures in
the range of $T_{\rm eff}\sim4000-4400$\,K (a similar 'bump' in
the $T_{\rm eff}$--color relations based on Pi98 colors can also
be seen in the $T_{\rm eff}$--$(V-I)$ and $T_{\rm eff}$--$(J-K)$
planes). It is quite possible, however, that the effective
temperatures of Pi98 giants are slightly overestimated in this
$T_{\rm eff}$ range (see Sect.~\ref{spectra} for details).

\begin{figure}[t]
\centering
\includegraphics[width=8.5cm] {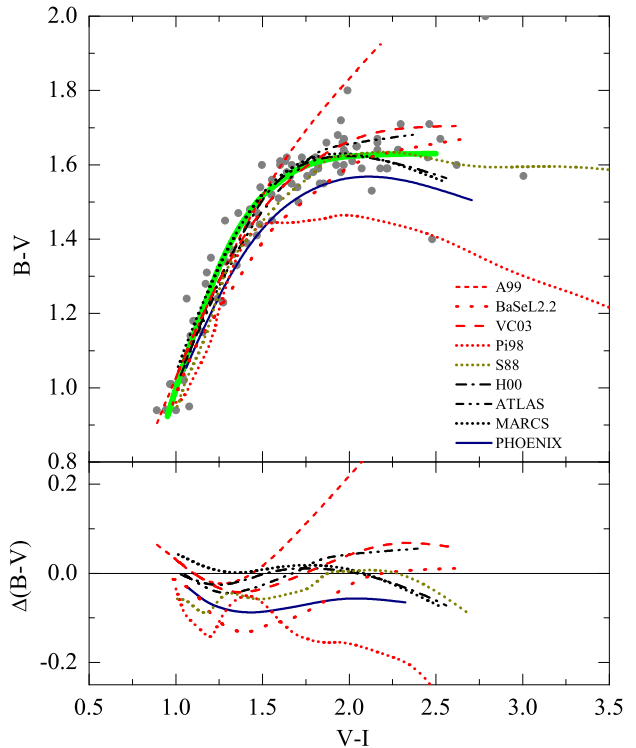}
\caption[]{(a) {\bf Top}: Empirical and theoretical color--color
relations in the $(B-V)$--$(V-I)$ plane. Filled circles are
late-type giants from Table~\ref{samplestars}. The thick green
line is a new empirical $(B-V)$--$(V-I)$ scale
(Table~\ref{empscale}; note that it is derived from the individual
best-fits in the $T_{\rm eff}$--$(B-V)$ and $T_{\rm eff}$--$(V-I)$
planes, i.e., it {\it is not} a best-fit in the $(B-V)$--$(V-I)$
plane). Several existing $T_{\rm eff}$--$(B-V)$ relations are also
shown. {\bf Bottom}: the difference between various
$(B-V)$--$(V-I)$ relations and the new empirical scale, $\Delta
T_{\rm eff}=T_{\rm eff}^{\rm other}-T_{\rm eff}^{\rm bestfit}$.
\label{CCplanes} }
\end{figure}

\begin{figure}[t]
\centering \addtocounter{figure}{-1}
\includegraphics[width=8.5cm] {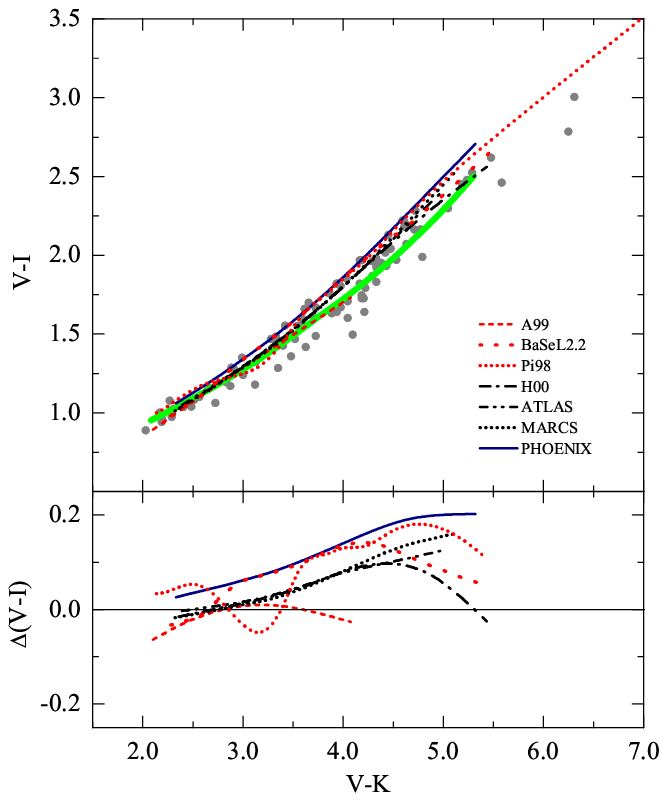}
\caption[]{(b) Same as Fig.~\ref{CCplanes}a but in the
$(V-I)$--$(V-K)$ plane.}
\end{figure}

Discrepancies are also small in the $T_{\rm eff}$--$(J-K)$ plane
(Fig.~\ref{TCplanes}d), where deviations do not exceed $\pm100$\,K
within $T_{\rm eff}=3500$--4500\,K (with the exception of the
scales based on Pi98 and BaSeL 2.2. colors). There is a hint
however, that our new empirical scale tends to predict slightly
higher effective temperatures for $(J-K)\sim0.6$--1.0 than other
existing scales. A slightly larger spread of different $T_{\rm
eff}$--color scales is seen in the $T_{\rm eff}$--$(V-I)$ plane,
though differences between the new empirical $T_{\rm
eff}$--$(V-I)$ scale and other existing relations generally are
within $\pm120$\,K (the $T_{\rm eff}$--color relations based on
Pi98, S88, and BaSeL 2.2 colors are somewhat more deviant).

The situation is more complex in the $T_{\rm eff}$--$(B-V)$ plane
(Fig.~\ref{TCplanes}a), where different scales start to diverge
below $\sim$4000\,K. These deviations are most obviously due to
inconsistencies in reproducing the `turn-off' towards the bluer
colors seen in the observed data at $(B-V)\sim1.65$ ($T_{\rm
eff}\sim3600$\,K) caused by the increasing absorption in TiO
bands. The majority of $T_{\rm eff}$--$(B-V)$ relations agree well
with the new empirical scale down to $\sim$3800\,K (or
$(B-V)\sim1.6$), the typical deviations being well within
$\pm100$\,K. The deviations are even smaller in case of VC03 scale
($\pm50$\,K down to $\sim$3900\,K). It should be noted in this
respect that theoretical isochrones of \citet{BV01} transformed to
observational planes using VC03 $T_{\rm eff}$--color relations are
in excellent agreement with the observed CMDs of Galactic globular
and open clusters, and field stars in the solar neighborhood
(VC03).

However, the agreement is considerably poorer in the case of
$T_{\rm eff}$--$(B-V)$ scales based on {\tt PHOENIX} and BaSeL 2.2
colors. The former starts to deviate at $\sim$4100\,K, while the
latter is discrepant by more than $\sim$100--200\,K within the
entire effective temperature range of interest. The scale based on
Pi98 colors shows significant deviations as well. These grow
especially large below $T_{\rm eff}\sim4000$\,K, where the Pi98
scale essentially fails to reproduce the observations and is at
odds with the other $T_{\rm eff}$--$(B-V)$ relations too. This
fact is rather disturbing, especially as the Pi98 scale has
difficulties in reproducing the observed colors of late-type
giants in other $T_{\rm eff}$--color planes too. Since the
spectral library of \citet{Pi98} is widely used in a variety of
astrophysical applications (e.g., the modelling of integrated
colors of stellar populations), these inconsistencies should be
properly taken into account. Finally, we should stress that none
of the existing $T_{\rm eff}$--$(B-V)$ scales reproduces trends in
the observed data correctly below $\sim$3800\,K. The only
exception in this sense is perhaps the scale based on S88 colors,
though it predicts the turn-off towards the bluer $B-V$ at
slightly higher effective temperatures than hinted from the
observations of late-type giants in our sample.

\begin{figure}[t]
\centering \addtocounter{figure}{-1}
\includegraphics[width=8.5cm] {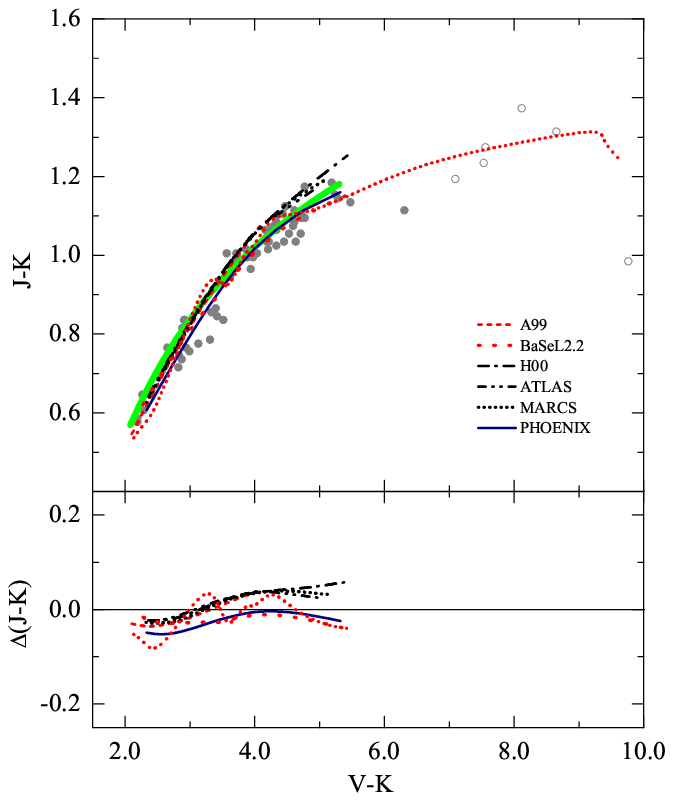}
\caption[]{(c) Same as Fig.~\ref{CCplanes}a but in the
$(J-K)$--$(V-K)$ plane. Stars from the sample of \citet{P98} are
highlighted as open circles.}
\end{figure}

It should be remembered that $T_{\rm eff}$--$\log g$--color scales
employing synthetic broad-band colors ({\tt PHOENIX}, {\tt MARCS},
{\tt ATLAS}) are indeed sensitive to the $T_{\rm eff}$--$\log g$
scale used, especially at lower effective temperatures. A shift in
gravity of $\Delta\log g= \pm0.2$ at $T_{\rm eff}=3500$\,K would
produce a shift $\Delta (B-V) \sim \pm 0.04$ and $\Delta(V-I )\sim
\Delta(V-K) \sim \pm0.03$ (redder colors for lower gravities),
with smaller differences for other colors. This would correspond
to differences of $\Delta T_{\rm eff} \sim50$\,K for $B-V$ and
$\Delta T_{\rm eff} \la 10$\,K for $V-I$ and $V-K$. Note that
these differences will be generally smaller at higher effective
temperatures, because of the weaker sensitivity of the emerging
spectral flux on gravity. While these effects may be important in
the $T_{\rm eff}$--$(B-V)$ plane, they are indeed too small to
influence the differences between different $T_{\rm eff}$--color
relations involving other photometric colors.

It is interesting to note, that synthetic colors of {\tt PHOENIX},
{\tt MARCS}, {\tt ATLAS} and H00 agree to within $\Delta T_{\rm
eff} \sim100$\,K over a large range of effective temperatures,
despite the fact that {\tt PHOENIX} models assume spherical
geometry while colors of {\tt MARCS}, {\tt ATLAS} and H00 are
calculated using plane-parallel model atmospheres. All of them
also show a reasonably good agreement with the observed colors of
late-type giants (to about $\sim150$\,K).

\begin{figure*}[t]
\center
\includegraphics[width=16cm]{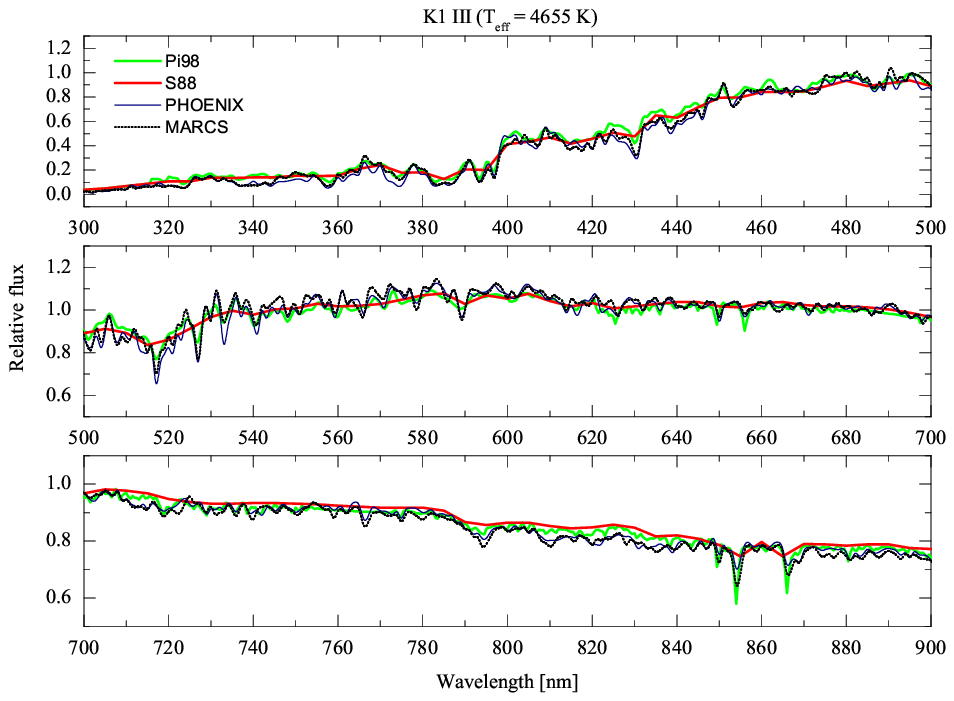}
\caption[]{~Comparison of observed spectra from the spectral
libraries of S88 and Pi98 with those calculated using {\tt
PHOENIX} and {\tt MARCS} model atmospheres, for K1~III giant
($T_{\rm eff}=4655$\,K, $\FeoH=0.0$; the metallicity of Pi98
spectrum is $\FeoH=0.09$). Surface gravity in theoretical models
was assigned according to the $T_{\rm eff}$--$\log g$ scale of H00
($\log g = 2.60$). Note the different y-scale in different panels.
\label{K1III}}
\end{figure*}

\begin{figure*}[t]
\center
\includegraphics[width=16cm]{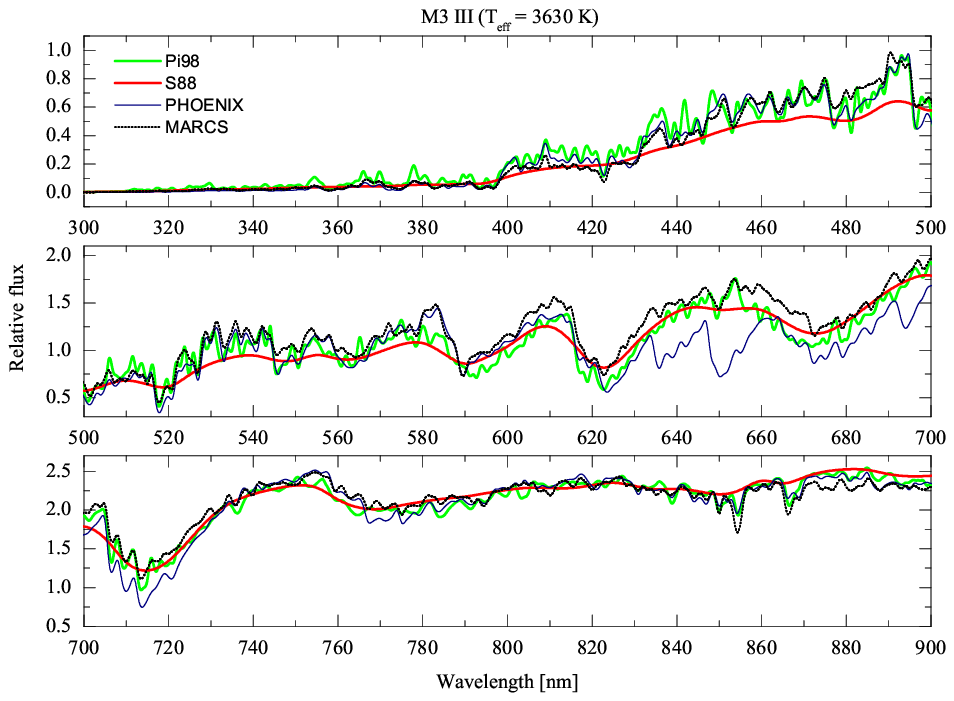}
\caption[]{~Same as in Fig.~\ref{K1III} but for a M3~III giant
($T_{\rm eff}=3630$\,K, $\log g = 0.62$, $\FeoH=0.0$).
\label{M3III}}
\end{figure*}

\subsection{Comparison of color--color relations}

The comparison of different color--color relations is given in
Figs.~\ref{CCplanes}a--c. Empirical color--color relations shown
there are constructed from $T_{\rm eff}$--color relations given in
Table~\ref{empscale}, thus they {\it do not} represent the
best-fits to the observed data in color--color diagrams.

The most complex behavior is seen in the $(B-V)$--$(V-I)$ plane.
While the agreement between different color--color scales is
generally rather good at higher effective temperatures (with the
exception of scale based on BaSeL 2.2 colors), discrepancies
become larger at lower $T_{\rm eff}$, which is to a large extent
determined by the discrepancies in the $T_{\rm eff}$--$(B-V)$
plane. For instance, the scale of A99 deviates rapidly beyond
$(B-V, V-I)~\sim~(1.5,1.5)$, and the Pi98 scale is strongly
deviant beyond $(B-V, V-I)~\sim~(1.4,1.4)$, as both of them fail
to reproduce the `turn-off' in the $T_{\rm eff}$--$(B-V)$ plane.
All three scales that are based on synthetic colors
(Tables~\ref{sempscale1}--\ref{sempscale2}) display slightly
differing trends, though they all agree with the observed colors
of late-type giants to within $\pm0.1$\,mag (or less) in $B-V$.
Note, however, that the scale based on {\tt PHOENIX} colors is
slightly too blue in $B-V$ throughout the entire effective
temperature range, which is again a consequence of the
discrepancies in the $T_{\rm eff}$--$(B-V)$ plane. Interestingly,
the scale based on colors calculated from the observed spectra of
S88 shows a remarkably good agreement with the observations in the
entire color--color range.

There seem to be some inconsistencies between observed and
synthetic $V-I$ colors within the entire effective temperature
range in the $T_{\rm eff}$--$(V-I)$ and $(V-I)$--$(V-K)$ planes,
as scales based on synthetic colors develop a systematical shift
towards the redder $(V-I)$ for a given value of $(V-K)$. This
effect is most pronounced for the {\tt PHOENIX} colors, largely
due to a somewhat discrepant trend in the $T_{\rm eff}$--$(V-I)$
plane. BaSeL 2.2 and Pi98 colors follow similar pattern. It is
rather unlikely though that these differences are due to the
systematical differences between the $I$-band magnitudes in the
Two-Micron Sky Survey and standard Kron systems
(Sect.~\ref{obscolors}), since the largest systematical
discrepancies are seen for the bluer colors and they gradually
diminish for the redder, while the differences in
Fig.~\ref{CCplanes}b are largest for the reddest stars. The good
agreement of the A99 scale with observed trends is somewhat
misleading. Since this scale is systematically shifted towards the
lower $T_{\rm eff}$ by a comparable amount both in the $T_{\rm
eff}$--$(V-I)$ and $T_{\rm eff}$--$(V-K)$ planes, these
differences compensate for each other and produce a good agreement
of the A99 scale with observations in the $(V-I)$--$(V-K)$ plane.

Excellent agreement is seen again in color--color diagrams
involving near-infrared colors; differences between various
color--color relations are typically within $\Delta(J-K)=0.1$ in
the $(J-K)$--$(V-K)$ plane. This fact is rather remarkable,
especially if taken into account that spectra of late-type giants
are influenced by a number of strong molecular bands (TiO, CO,
${\rm H_2O}$, VO, CO, etc.) in this wide wavelength range.

\subsection{Comparison of observed and synthetic spectra\label{spectra}}

\begin{figure*}[p]
\center
\includegraphics[width=18cm]{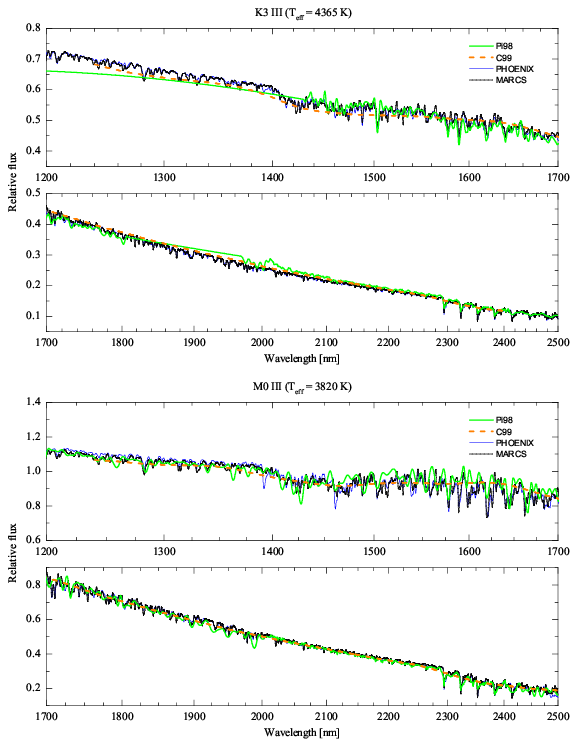}
\caption[]{~Comparison of the near-infrared observed spectra from
the spectral libraries of Pi98 and C99 with those calculated using
{\tt PHOENIX} and {\tt MARCS} model atmospheres, for K3~III
($T_{\rm eff}=4365$\,K, top) and M0~III ($T_{\rm eff}=3820$\,K,
bottom) giants. To make the comparison easier, spectra are
renormalized to yield the same total flux in the wavelength
interval covered by the C99 spectrum. All spectra are of solar
metallicity except K3~III of Pi98 ($\FeoH=-0.02$). Surface gravity
in theoretical models was assigned according to the $T_{\rm
eff}$--$\log g$ scale of H00 ($\log g = 2.01$ and $\log g=0.97$
for K3~III and M0~III, respectively). \label{IRspectra}}
\end{figure*}

\begin{figure*}[t]
\center
\includegraphics[width=18cm]{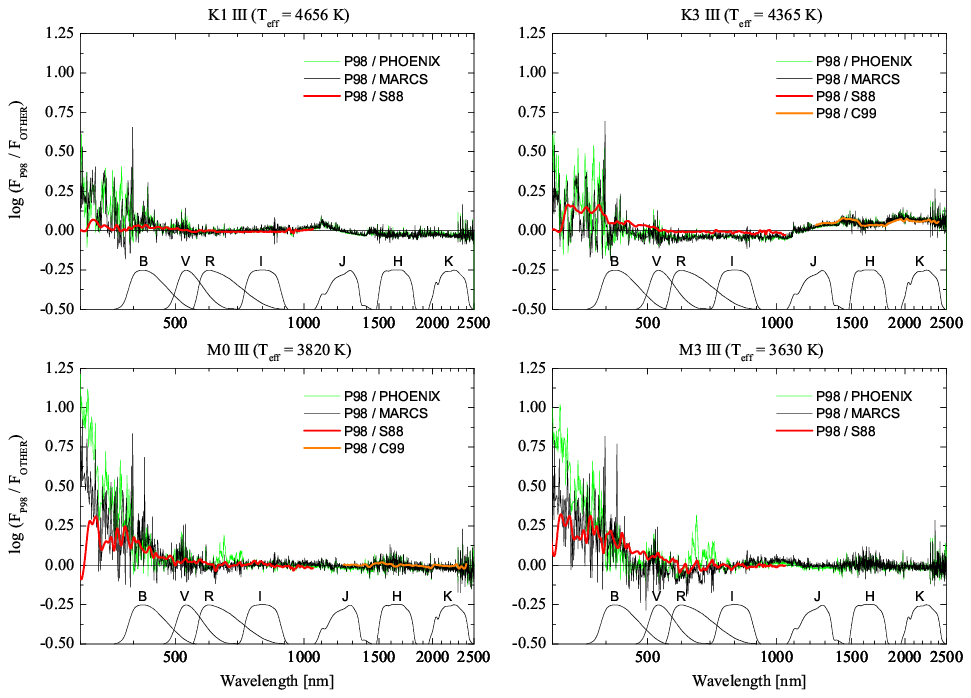}
\caption[]{~Spectral flux ratios of Pi98 spectra over the other
observed (S88, C99) and synthetic ({\tt PHOENIX}, {\tt MARCS})
spectra. \label{Fluxratios}}
\end{figure*}

Perhaps the most direct way to clarify what causes the differences
between the observed photometric colors of late-type giants and
synthetic colors calculated with different stellar model
atmospheres is by making a comparison of observed and synthetic
spectra.

Examples of the observed optical spectra from the spectral
libraries of S88 and Pi98, together with those calculated with the
{\tt PHOENIX} and {\tt MARCS} model atmospheres are shown in
Figs.~\ref{K1III}-\ref{M3III}, for late-type giants of spectral
classes K1~III ($T_{\rm eff}=4655$\,K, according to the spectral
class--effective temperature relation used in Pi98) and M3~III
($T_{\rm eff}=3630$\,K). The observed spectra of Pi98 correspond
to the metallicities of $\FeoH=0.09$ and $\FeoH=0.0$ for the
spectral types K1~III and M3~III, S88 spectra are of solar
metallicity. Synthetic spectra are all calculated at solar
metallicity, for the effective temperatures corresponding to the
two spectral classes and with gravities assigned according to the
$T_{\rm eff}$--$\log g$ relation of H00 ($\log g=2.60$ and $\log
g=0.62$, for K1~III and M3~III, respectively). Synthetic spectra
are degraded to a resolution of 1\,nm (which corresponds to a
typical resolution in the Pi98 library), and then are normalized
in such a way that they yield the same total flux as the
corresponding Pi98 spectrum in the wavelength interval covered by
the latter. S88 spectra are normalized to yield the same flux as
Pi98 spectra in the wavelength interval covered by S88 spectra.
Note that resolution of S88 spectra ($\sim5$\,nm) is certainly too
low for making a detailed comparison on the level of individual
spectral lines/bands. Nevertheless, they are still useful for the
comparison of general trends, especially given the fact that
$T_{\rm eff}$--color relations based on photometric colors of
late-type giants in S88 are in excellent agreement with the
observations and other $T_{\rm eff}$--color scales in the $T_{\rm
eff}$--$(B-V)$ plane. Similarly, Fig.~\ref{IRspectra} illustrates
the situation in the infrared part of the spectrum, for the
spectral types of K3~III and M0~III. Here, we also show infrared
spectra of two late type giants from the spectral library of C99:
$\alpha$\,Hya \citep[K3~III; the original spectrum taken
from][]{C95} and $\beta$\,And \citep[M0~III;][]{C95}. Note that
all spectra in Fig.~\ref{IRspectra} are normalized in such a way
that they yield the same total flux over the wavelength interval
covered by the C99 spectrum (1240--2400\,nm in case of K3~III and
1240--2500\,nm in case of M0~III). Finally, Fig.~\ref{Fluxratios}
shows the flux ratios of Pi98 spectra over the other observed
(S88, C99) and synthetic ({\tt PHOENIX}, {\tt MARCS}) spectra (the
y-scale in Fig.~\ref{Fluxratios} is a logarithm of the spectral
flux ratios, thus if multiplied by 2.5 it will provide the
difference in monochromatic magnitudes at a given wavelength).

Generally, the agreement between the observed and synthetic
spectra is rather good at $\sim500-900$\,nm, especially at higher
effective temperatures. However, there are numerous small
differences in reproducing the strength of individual lines/bands
and the continuum level, first of all due to inadequate knowledge
of atomic and especially molecular opacities (note, for example,
the difficulties in reproducing the observed flux at around
$\sim420-440$\,nm, which is determined by the strength of CH
G-band and numerous atomic lines, Fe~I, Ti~I, Ti~II, Ca~I; or the
differences in the continuum level at $\sim810-840$\,nm --
Figs.~\ref{K1III}, \ref{Fluxratios}a). This is evident both with
{\tt PHOENIX} and {\tt MARCS} spectra, though the latter seem to
be in a slightly better agreement with observations. These
discrepancies grow larger at lower effective temperatures, with
differences in reproducing molecular band strengths becoming
increasingly important (e.g., note the difficulties in reproducing
TiO bands in the range of $584-800$\,nm, with bandheads at 584.7,
615.9, 665.1, and 758.9\,nm).

The differences between observed and synthetic spectra, however,
become progressively larger at shorter wavelengths ($\la500$\,nm),
and they also tend to increase with the decreasing effective
temperature. It should be noted though, that this spectral range
is rather difficult to model theoretically since even minor
changes in the model parameters have a strong influence on the
emerging spectrum at $\lambda\la500$\,nm. On the observational
side, the S/N ratio in the observed spectra of late-type giants is
low at short wavelengths (because of the low spectral flux), thus
precise absolute flux calibration is indeed difficult to achieve.
Note, for instance, the differences in the spectral flux between
the spectral libraries of S88 and Pi98 at $\lambda\la500$\,nm,
which are responsible for their significantly different $B-V$
colors and thus for the different behaviour of the two $T_{\rm
eff}$--color scales in the $T_{\rm eff}$--$(B-V)$ plane.
Obviously, further improvements are urgently needed here, both in
terms of new observations and improved theoretical models, to
bridge the gap between the theoretical predictions and
observations in this wavelength range.

Note, however, that there are significant differences between the
observed K3~III spectrum of Pi98 and those calculated with {\tt
PHOENIX} and {\tt MARCS} model atmospheres, over the entire
wavelength range considered in this study (differences between the
Pi98 and S88 spectra are somewhat smaller because S88 spectrum was
employed in the derivation of the K3~III spectrum of Pi98). This
is clearly seen in Fig.~\ref{Fluxratios}b, which hints towards an
obvious residual trend in the range of $500-2500$\,nm. This
residual trend could be removed by assigning a somewhat lower
($\sim100$\,K) effective temperature to the `K3~III' spectrum of
Pi98 (situation with the K2~III and K4~III spectra of Pi98 is
qualitatively very similar). This would also solve the problem
with the 'bump' in the $T_{\rm eff}$--$(V-I)$ and $T_{\rm
eff}$--$(V-K)$ relations based on the Pi98 colors (and $T_{\rm
eff}$--$(V-I)$ relation based on the colors of S88) discussed in
Sect.~\ref{Teff-color}.

The agreement between the observed and synthetic spectra is
generally rather good at the near-infrared wavelengths too,
especially given the fact that this wavelength range is strongly
affected by various molecular bands (Fig.~\ref{IRspectra}).
Indeed, differences are abundant too, e.g., those in reproducing
the strength of water and (to a smaller extent) CO bands at
$\sim1.3-1.8$\,$\mu$m, which is well noticeable at lower effective
temperatures (note that observed CO bands are rather well
reproduced at $\sim2.3-2.5$\,$\mu$m, even for M0~III).

Thus in overall, despite the fact that both {\tt PHOENIX} and {\tt
MARCS} models are able to reproduce the general trends in the
observed spectra rather well (which is also reflected in the
relatively good agreement between the synthetic colors and
observations), there are numerous differences both in the
predicted strengths of individual lines/bands and/or continuum
level. In many cases these differences are large enough to cause
noticeable changes even in the broad-band photometric colors.

\section{Summary and conclusions}

We have calculated a new {\tt PHOENIX} grid of synthetic
broad-band photometric colors, which covers the effective
temperatures $T_{\rm eff}=3000\dots 5000$\,K, gravities $\log
g=-0.5\dots{+3.5}$, and metallicities $\MoH=+0.5\dots{-4.0}$. Our
analysis shows that synthetic colors of late-type giants are
noticeably influenced by a number of different model parameters.
The influence of various molecular bands is strong in all studied
$T_{\rm eff}$--color and color--color planes below $T_{\rm
eff}\sim4000$\,K. The effects of TiO are most prominent in the
optical wavelength range, while H$_2$O strongly affects the
near-infrared colors. All photometric colors are strongly
influenced by the effects of gravity at lower effective
temperatures (typically, below $T_{\rm eff}\sim3600$\,K). This is
associated with the fact that molecule formation is generally more
efficient at lower gravities, since the outer atmospheric layers
are getting more extended and cooler. The influence of gravity is
strongest in the $T_{\rm eff}$--$(B-V)$ plane, which is due to the
fact that $B-V$ color is strongly affected by different molecular
lines. The effect of the microturbulent velocity is well
noticeable at all effective temperatures, the difference in
certain photometric colors may reach $\sim0.2$\,mag in response to
a change in microtubulent velocity from $\xi=1.0$ to
$5.0$\,km\,$s^{-1}$ (the effect is marginally smaller at higher
gravities). The influence of stellar mass on photometric colors is
generally small but non-negligible at lower effective
temperatures, which is related to differences in atmospheric
structures for different stellar masses, with higher mass model
atmospheres being marginally hotter in the outer (optically thin)
layers.

We find that also convection may influence photometric colors in a
non-negligible way. The difference between synthetic colors
calculated with a fully time-dependent 3D hydrodynamical model
atmosphere and those obtained with the conventional 1D model may
reach up to several tenths of a magnitude in certain photometric
colors (e.g., $V-K$), equivalent to a shift in effective
temperature of up to $\sim$70\,K. This fact is rather interesting,
since standard 1D models predict that convection is restricted to
the optically thick layers, and thus has no influence on the
atmospheric structure and spectroscopic or photometric properties
(which we confirm by finding no differences in photometric colors
calculated with 1D model atmospheres using different mixing length
parameter). In contrast, our full hydrodynamical modeling predicts
an intense convective overshoot from the deeper interiors in to
the outer atmospheric layers, which has a non-negligible influence
on the photometric colors too.

To compare the new synthetic photometric colors with observations
of late-type giants, we derive a new $T_{\rm eff}$--$\log
g$--color relation based on observed quantities of a homogeneous
sample of late-type giants in the solar neighborhood, with
effective temperatures available from interferometry and surface
gravities selected according to the $T_{\rm eff}$--$\log g$
relation of H00. Given the typical errors of interferometrically
derived $T_{\rm eff}$, systematical effects, etc., the internal
precision of these empirical $T_{\rm eff}$--color relations is
limited to about $\pm150$\,K. We also provide three additional
semi-empirical scales based on the $T_{\rm eff}$--$\log g$
relation of H00 and synthetic colors of {\tt PHOENIX}, {\tt MARCS}
and {\tt ATLAS} (the new {\tt MARCS} spectra/colors were kindly
provided by B. Plez, {\tt ATLAS} colors were taken from Castelli
\& Kurucz 2003).

Generally, the new $T_{\rm eff}$--color relations based on
synthetic colors are in good agreement with observations of
late-type giants (to about $\pm150$\,K or better). The deviations
between the relations based on different stellar atmosphere models
are small, typically within $\pm100$\,K or less, with slightly
larger discrepancies seen in the $T_{\rm eff}$--$(V-I)$ plane. The
situation is more complex in the $T_{\rm eff}$--$(B-V)$ plane.
While the agreement between observed and synthetic colors is
rather good at higher effective temperatures, all scales tend to
disagree below $\sim$3800\,K, which is related to difficulties in
reproducing the `turn-off' towards the bluer colors seen in the
observed data at $T_{\rm eff}\sim3600$\,K (caused by the
increasing strength of TiO bands at lower effective temperatures).

The agreement between different color--color scales in the
$(B-V)$--$(V-I)$ plane is rather good for $(B-V, V-I) \la
(1.5,1.5)$ and degrades rapidly for the redder colors, which is
defined essentially by the trends exhibited by different scales in
the $T_{\rm eff}$--$(B-V)$ plane. There are some discrepancies in
the $(V-I)$--$(V-K)$ plane too, where synthetic colors tend to
become redder in $V-I$ at lower effective temperatures. Excellent
agreement is seen in the $(J-K)$--$(V-K)$ plane, where differences
between the trends of different color--color scales do not exceed
$\pm0.05$ in $J-K$.

Finally, we make a brief comparison of the observed and synthetic
spectra of late-type giants at several effective temperatures
typical to late-type giants, both at optical and near-infrared
wavelengths. While in general the agreement between the
observations and theoretical predictions is reasonably good, the
differences are abundant too, both in reproducing the strengths of
individual spectral lines/bands (especially - molecular bands,
e.g., TiO, H$_2$O, CO and so forth) and the continuum level. These
differences are large enough to produce noticeable discrepancies
at the level of broad-band photometric colors.

\acknowledgements

We are grateful to Bertrand Plez (GRAAL, Universit\'{e}
Montpellier) for calculating {\tt MARCS} grid of synthetic
spectra, and numerous comments and discussions. We thank Glenn
Wahlgren (Lund Observatory) for a careful reading of the
manuscript and his valuable comments and suggestions. We are also
indebted to the referee, Michael Bessel, for his very constructive
and valuable suggestions that helped to improve the paper
considerably. AK acknowledges support from the Wenner-Gren
Foundations. This work was supported in part by grant-in-aids for
Scientific Research (C) and for International Scientific Research
(Joint Research) from the Ministry of Education, Science, Sports
and Culture in Japan, and by a Grant of the Lithuanian State
Science and Studies Foundation. It was supported in part by NSF
grants AST-9720704 and AST-0086246, NASA grants NAG5-8425,
NAG5-9222, as well as NASA/JPL grant 961582 to the University of
Georgia.  This work was also supported in part by the P\^ole
Scientifique de Mod\'elisation Num\'erique at ENS-Lyon. Some of
the calculations presented in this paper were performed on the IBM
pSeries 690 of the Norddeutscher Verbund f\"ur Hoch- und
H\"ochstleistungsrechnen (HLRN), on the IBM SP ``seaborg'' of the
NERSC, with support from the DoE, and on the IBM SP ``Blue
Horizon'' of the San Diego Supercomputer Center (SDSC), with
support from the National Science Foundation.  We thank all these
institutions for a generous allocation of computer time. This
research has also made use of the SIMBAD and VizieR databases,
operated by the CDS, Strasbourg, France.

\bibliographystyle{aa}

\end{document}